%% file: 0-Main.tex
\documentclass[pdflatex,sn-apa]{sn-jnl}

\usepackage{graphicx,calc}%
\usepackage{graphicx}%
\usepackage{array}%
\usepackage{multirow}%
\usepackage{amsmath,amssymb,amsfonts}%
\usepackage{amsthm}%
\usepackage{mathrsfs}%
\usepackage[title]{appendix}%
\usepackage{xcolor}%
\usepackage{textcomp}%
\usepackage{manyfoot}%
\usepackage{booktabs}%
\usepackage{algorithm}%
\usepackage{algorithmicx}%
\usepackage{algpseudocode}%
\usepackage{listings}%
\usepackage{wrapfig}

\usepackage[utf8]{inputenc}%
\usepackage[T2A]{fontenc}%
\usepackage{caption} 
\usepackage[misc,geometry]{ifsym}
\usepackage{subcaption}
\usepackage[font=small]{caption, subcaption}
\usepackage[section]{placeins}
\usepackage{flushend}
\usepackage{pdflscape}
\usepackage{adjustbox}
\usepackage[english]{babel}
\usepackage{colortbl}

\theoremstyle{thmstyleone}%
\theoremstyle{thmstyletwo}%

\theoremstyle{thmstylethree}%

\begin{document}

\title[Article Title]{\Large Fairness in Computational Innovations: Identifying Bias in Substance Use Treatment Length of Stay Prediction Models with Policy Implications}


\author{\fnm{Ugur} \sur{Kursuncu}}\email{ugur@gsu.edu}

\author{\fnm{Aaron} \sur{Baird}}\email{abaird@gsu.edu}

\author{\fnm{Yusen} \sur{Xia}}\email{ysxia@gsu.edu}

\affil{\orgdiv{\centering Institute for Insight, J. Mack Robinson College of Business, \\ 
Georgia State University}, \state{Atlanta, GA}, \country{USA}}




\abstract{Predictive machine learning (ML) models are computational innovations that can enhance medical decision-making, including aiding in determining optimal timing for discharging patients. However, societal biases can be encoded into such models, raising concerns about inadvertently affecting health outcomes for disadvantaged groups. This issue is particularly pressing in the context of substance use disorder (SUD) treatment, where biases in predictive models could significantly impact the recovery of highly vulnerable patients. In this study, we focus on the development and assessment of ML models designed to predict the length of stay (LOS) for both inpatients (i.e., residential) and outpatients undergoing SUD treatment. We utilize the Treatment Episode Data Set for Discharges (TEDS-D) from the Substance Abuse and Mental Health Services Administration (SAMHSA). Through the lenses of distributive justice and socio-relational fairness, we assess our models for bias across variables related to demographics (e.g., race) as well as medical (e.g., diagnosis) and financial conditions (e.g., insurance). We find that race, US geographic region, type of substance used, diagnosis, and payment source for treatment are primary indicators of unfairness. From a policy perspective, we provide bias mitigation strategies to achieve fair outcomes. We discuss the implications of these findings for medical decision-making and health equity. We ultimately seek to contribute to the innovation and policy-making literature by seeking to advance the broader objectives of social justice when applying computational innovations in health care.}


\maketitle

\section{Introduction}
\label{sec:intro}

\input{1-Introduction}

\section{Background and Theory}
\label{sec:backgrounTheory}
\input{2-Background-Theory}

\section{Methods}
\label{sec:methods}
\input{3-Methods}

\section{Results}
\label{sec:results}
\input{4-Results}

\section{Discussion}
\label{sec:discusssion}
\input{5-Discussion}

\section{Conclusion}
\label{sec:conclusion}

\input{6-Conclusion}

\bibliography{sn-bibliography}

\clearpage

\input{7-Appendix}

\end{document}

%% file: 1-Introduction.tex
Predictive Machine Learning (ML) models are increasingly used in medical decision-making \citep{shahid2019neuralnetworks,secinaro2021aihealthcare,bertl2023ai,khosravi2024decisionmaking}. Unfortunately, however, social biases can be inadvertently encoded into such models \citep{chouldechova2020snapshot,ganju2020role,li2017feature,soenksen2022framework}. A recent study revealed that an ML model used in US hospitals to manage access to resources was biased against Black patients, causing additional medical or financial distress to underserved groups already suffering from societal bias \citep{obermeyer2019dissecting}. Hence, revealing potential bias in ML models has become a key focus for researchers and policymakers \citep{hall2019taxonomy,landers2023auditing,toledano2024ethics,wiersma2023ethics}.

Socially fair treatment is imperative for Substance Use Disorder (SUD) patients undergoing rehabilitation, as any potential unfairness may exacerbate pre-existing medical and financial conditions. A particularly critical decision is how long a patient needs to stay in a facility or a rehabilitation program for observation and treatment (i.e., length-of-stay, LOS). In this context, LOS is a critical factor for a patient to receive the most positive health care experience and optimal outcomes\footnote{\texttt{https://data.oecd.org/healthcare/length-of-hospital-stay.htm}} \citep{stone2022los}. More specifically, longer LOS in a SUD treatment facility or program often leads to better mental health care outcomes in terms of medical well-being of the patient \citep{mennis2019treatment,baird2022machine}. Keeping a SUD patient for shorter LOS than necessary in a facility or program may lead to negative outcomes in terms of medical well-being of the patient \citep{rotter2010pathways}. While prior studies have focused on disparities in SUD wait times \citep{kong2022opioid} and treatment completion \citep{baird2022machine}, LOS with respect to fairness and disparities for SUD treatment has not been fully examined. 

Thus, in this work, consistent with the urgent need for fairness in predictive ML model development for decision-making in health care \citep{lee2021mental,shin2019algorithmic,nordling2019fairer,paulus2020predictably,rajkomar2018fairness}, we investigate the fairness of ML models developed for prediction of LOS in SUD treatment. Our objective is to identify social subgroups that may be adversely impacted by ML model predictions, with implications for managing computational innovations and developing associated mitigation strategies and policy. We leverage the U.S. nationwide Treatment Episode Data Set for Discharges (TEDS-D) from the Substance Abuse and Mental Health Services Agency (SAMHSA), which provides discharge data from detoxification, ambulatory (i.e., outpatient), and residential (i.e., inpatient) SUD treatment programs. We develop ML models to predict LOS for SUD inpatients (i.e., those in residential treatment programs) and outpatients. We then assess these models for fairness based on variables related to the demographics, financial status, and medical conditions of the patients. We find that race, geographic region, type of substance used, Diagnostic and Statistical Manual of Mental Disorders (DSM) diagnosis, and payment source for treatment are the primary variables exhibiting higher levels of unfairness. 

In this context, LOS is a critical factor for a patient to receive the most positive health care experience and optimal outcomes \citep{stone2022los}. More specifically, longer LOS in an SUD treatment facility or program often leads to better mental health care outcomes in terms of medical well-being of the patient \citep{mennis2019treatment,baird2022machine}. On the other hand, keeping an SUD patient for shorter LOS than necessary in a facility or program may lead to negative outcomes in terms of medical well-being of the patient \citep{rotter2010pathways}. This negative impact may be exacerbated in society in the form of socially discriminative and unfair outcomes when an ML model is involved in decision-making. Prior studies focused on disparities in SUD wait times \citep{kong2022opioid} and treatment completion \citep{baird2022machine}, while surprisingly and to our knowledge, LOS with respect to fairness and disparities for SUD treatment has not been explored. Further, those diagnosed with SUD are often treated as in-patients within facilities and/or outpatients in SUD treatment programs that can range from ambulatory to intensive residential programs. Prior research tends to focus on either inpatient or outpatient but has yet to consider both in relation to fairness. Thus, in this paper, we focus on the LOS for patients being treated for SUD, and developed and assess models for both inpatient and outpatient treatment LOS.

Consequently, we ultimately seek to contribute in three ways, relative to the literature on distributive justice \citep{rajkomar2018fairness,rea2021unequal} and model fairness \citep{paulus2020predictably,samorani2021overbooked}. First, our findings shed light on potentially existing disparities in SUD treatment, especially where biases could be perpetuated through the use of ML. These findings expand our knowledge of how to responsibly manage innovations \citep{pandza2013strategic,owen2021organisational}. Second, we inform general knowledge and practices in fair and equitable mental health care management, focusing on multiple determinants of health outcomes in evaluating ML models for fairness. Such findings contribute to our understanding of the governance of AI/ML, especially when ethical issues are present \citep{goos2024governance,reddy2020governance}. Third, our approach can be adjusted and adapted for other critical problems in health care management and policymaking such as fair allocation of health care resources in other contexts, such as in emerging science \citep{kuhlmann2019tentative} and practical use of innovations \citep{devasconcelos2025failure}. By shedding light on potentially disadvantaged subgroups, we contribute to the current discourse and body of research on health equity and social justice \citep{rajkomar2018fairness,uddin2024fairness}. Overall, our study emphasizes the potential of data-driven approaches to inform public health strategies, ensuring they are grounded in equity \citep{ruijer2023equity}. 

In the sections that follow, we describe the background for this study in Section \ref{sec:backgrounTheory}, explain the methods and data in Section \ref{sec:methods}, provide results on the assessment of fairness in the models in Section \ref{sec:results}, and, in Section \ref{sec:discusssion}, we offer insights based on our findings, contributions, and opportunities for future work. 

%% file: 2-Background-Theory.tex
Fairness in AI is defined as “the absence of any prejudice or favoritism toward an individual or group based on their inherent or acquired characteristics…. [i.e.,] an unfair algorithm is one whose decisions are skewed toward a particular group of people” \citep[p.~115:2]{mehrabi2021bias}. Measures of fairness for AI, such as demographic parity, accuracy equity, and equality of opportunity, have been introduced and examined based on philosophical theories \citep{binns2018fairness}. Ensuring the responsible use of AI technologies by companies and society is an urgent challenge, given the rapid advancements in this technology and notable short-comings in high-stakes decision-making contexts \citep{fu2022unfair,dearteaga2022fairness,johnmathews2022reality}. In the following sections, we provide contextual background on critical concepts underpinning our focus on fairness, including distributive justice and social-relational aspects of health care as well as bias and fairness in AI models.

\subsection{Distributive Justice Theory for (Mental) Health Care}
Distributive justice theory addresses the equitable allocation of resources and opportunities across different groups in society \citep{rawls1971justice}. In his seminal philosophical work, “A Theory of Justice” (1971), John Rawls argues for a distributive justice-based system where social and economic inequalities are arranged to benefit the least advantaged members of society \citep{rawls1971justice}. He argues that just institutions should not make arbitrary distinctions between people and should ensure a fair balance of advantages in social cooperation. Thus, justice is defined by principles that guide the fair allocation of rights, duties, and social advantages. In extending this work and applying it to the health care context, \cite{rajkomar2018fairness} identifies three distributive justice approaches to assess fairness in health care ML models: (1) Equal patient outcomes, where models produce similar health outcomes for all groups; (2) Equal ML performance, ensuring that models perform equally well for all groups, based on metrics like accuracy, sensitivity, specificity (i.e., equalized odds), and positive predictive value (i.e., selection rate); and (3) Equal allocation, where ML decisions lead to a proportional allocation of resources, evaluated using demographic parity. 
\cite{giovanola2023fairness} further argue that fairness in health care ML systems should go beyond addressing discrimination and ensuring equal distribution. Discrimination is often viewed as unfair treatment rooted in a failure to recognize the equal moral worth of individuals \citep{dworkin2002sovereign}. Addressing discrimination requires ethical reasoning and a deeper understanding of fairness as an ethical value \citep{mccradden2020ethical}. From a socio-relational perspective, discrimination is seen as degrading and demeaning, preventing mutual recognition of equality \citep{anderson1999point,scheffler2017egalitarianism}. Thus, we include socio-relational considerations into our evaluation. Research shows that inherent biases in ML models potentially arising from disparities are often linked to the social determinants of health (SDOH), such as economic stability, education, healthcare access, and social support \citep{galea2007poverty,tolliver2010earlychildhood,yearby2022justice}. This aspect emphasizes not only equitable distribution of resources, but also promotion of non-discrimination, equal respect for individuals, and the creation of health care structures that affirm the moral worth of all patients, regardless of their socio-economic background. 

This dual perspective of distributive justice and socio-relational considerations acknowledges that fairness in health care extends beyond simply addressing disparities in resource allocation by shedding light on the societal conditions that contribute to unequal health care outcomes. Thus, fairness as an ethical value encompasses both distributive justice—ensuring equitable resource distribution—and socio-relational fairness—upholding the equal moral worth of all patients, regardless of their social or economic circumstances.

\subsection{AI Fairness in Health Care and SUD Treatment}
Fair access to health care resources is essential for maintaining and improving the health of an entire population, as opposed to only specific subgroups \citep{emanuel2020allocation}. While fairness and equity in health care can be debated from various perspectives \citep{olsen2011equity}, our study does not consider fairness as absolute equality. Following economic \citep{olsen2011equity}, and ethical \citep{chen2021ethical,morley2020ethics} perspectives on health care resource allocation, we acknowledge acceptable differences in resource use between subgroups based on medical needs. However, differences in resource use within groups with similar health needs are especially concerning, when unrelated to health factors.

While prior work has made several advances in understanding how to accurately make clinical predictions, fairness is not always a central consideration. For instance, \cite{song2015queue} improved LOS in emergency departments using a dedicated queuing system. \cite{belderrar2020application} identified $14$ common predictive factors influencing hospital LOS in intensive care units. \cite{kalgotra2021hospital} developed predictive models to predict the expected LOS at the time of admission, achieving the lowest average mean absolute percent error, by combining network science and deep learning methods. Yet, fairness was not a central consideration in these works. Further, protected group information (e.g., race, gender) is not always directly observed in the data \citep{chen2021ethical}. Researchers have proposed optimization-based methods to estimate the range of potential disparities in predicting protected class membership from proxy variables using auxiliary data, such as census data \citep{kallus2022algorithmic}. But a significant challenge remains in addressing disparities as potentially biased historical data may lead to algorithmic bias. Algorithms that do not correct for biases will likely perpetuate them \citep{wiens2020bias,parikh2019bias}.

For patients with SUDs, fairness in AI applications is especially crucial. For instance, biases in predicting the LOS for SUD treatment can lead to inadequate care for subgroups, such as those of a certain race or who do not have insurance, further worsening health disparities. In prior SUD-related research, novel ML methods have been applied to various aspects of SUD-related care, including outcome prediction \citep{nasir2021outcome}, treatment referrals \citep{afshar2022smartai}, and natural language analysis of EHR data \citep{riddick2022nlp} and social media data \citep{lokala2022drug,kursuncu2018s} to identify patients at risk of SUD. Yet, while these contributions are valuable, identifying and acknowledging such biases is crucial, as the discharge data used to train these algorithms likely contains inherent biases, reflecting known disparities. Disparities in SUD treatment wait times and treatment completion rates, such as those associated with race and ethnicity are well-documented \citep{baird2022machine,baird2023determinants,kong2022opioid,delphin2012racial}. Similarly, demographic disparities are prevalent in the LOS for various health care admissions, including general hospital inpatient admissions \citep{ghosh2022diversity,tan2019diabetic}. Despite these advancements, disparities specific to LOS for SUD treatment remain underexplored in the literature. This is even though SUD-related LOS is often longer and more resource-intensive than general admissions, especially when comorbidities such as other mental health disorders, are present \citep{ndanga2019hospitalization}. Hence, in contrast to prior research, our study focuses on LOS as the target variable within SUD treatment, revealing potential biases in ML models trained using these public datasets.

\begin{figure}[h!]
    \centering
    \includegraphics[width=\linewidth]{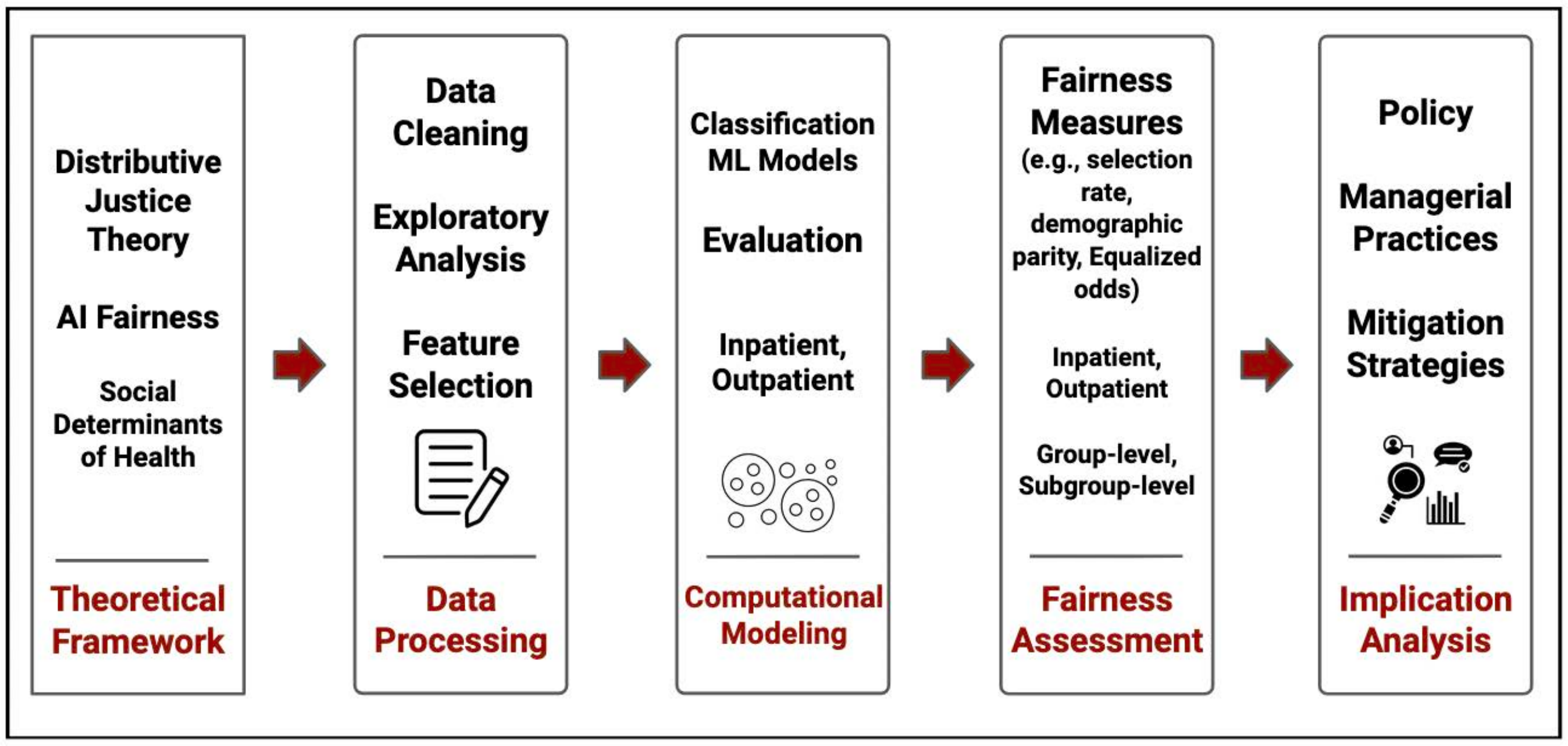}  
    \caption{Our approach, grounded in distributive justice theory, for developing ML models for predicting LOS, and assessing them for fairness based on two levels; inpatient and outpatient, group-level and subgroup-level.}
    \label{fig:diagram}
\end{figure}

%% file: 3-Methods.tex
Our methodological goals are two-fold: (i) to develop classification models that classify the LOS for patients into short-term (negative) or long-term (positive) SUD treatment, and (ii) to assess the fairness of these models as to whether they perform equally well for all social groups without any social discrimination; thus, revealing potential disparities affecting certain subgroups (see Figure \ref{fig:diagram}). We begin by describing the dataset used in this study and the selection of LOS as the target variable. Then, we provide details on our predictive modeling approach, including preprocessing of data, feature engineering and selection, and model building. We assess the fairness of the two best-performing predictive models by examining their performance across various subgroups characterized by demographic, medical, and financial factors. 

\subsection{Data}
We use the Treatment Episode Data Set: Discharges (TEDS-D)\footnote{\texttt{https://www.datafiles.samhsa.gov/dataset/teds-d-2019-ds0001-teds-d-2019-ds0001}} and we focus on the year of 2019. Examining this pre-pandemic period avoids the potential complications and new dynamics that might be introduced by the global pandemic, allowing for a clearer analysis of the standard health care environment. We use inpatient (i.e., residential) and outpatient (i.e., ambulatory) LOS as our target variable (see Section \ref{sec:targetVariableLOS}). IRB approval, as an exempt study due to the de-identified nature of the data, was applied for and received.

\subsubsection{TEDS-D from SAMSHA}
\label{sec:tedsSamsha}
SUD treatment facilities and programs across the U.S. are connected to an online data collection system administered by SAMSHA. These facilities and programs provide annual summary data for each patient admission and discharge to this central system. The de-identified discharge dataset, known as TEDS-D, is made publicly available and contains demographic data (e.g., race, age, gender), type of service at discharge, LOS, discharge reasons or discontinuation of service, geographical region, and substance use information (e.g., substances used, route of use, frequency of use). For the year 2019, TEDS-D includes $1,722,503$ patient records with $76$ variables. We removed seven variables (\texttt{FREQ3\_D}, \texttt{DETCRIM}, \texttt{FREQ3}, \texttt{FRSTUSE3}, \texttt{ROUTE3}, \texttt{DETNLF\_D}, \texttt{DETNLF}) with missing value ratios greater than $70\%$. Missing values (i.e., unknown or not provided by the patient) were treated as a separate category along with other categorical variables since the presence or the absence of sensitive or personal data might represent aspects of patient behavior or circumstances. We also removed record-specific variables that do not contribute to model training (e.g., \texttt{CASEID}, which is the ID number of the discharge). Given the distinct characteristics of inpatient and outpatient treatments, we split the dataset based on the \texttt{SERVICES\_D} variable which categorizes services from 1 to 8. Categories 3-5 indicate inpatient care (i.e., rehab/residential hospital non-detox, rehab/residential short term 30 days or fewer, rehab/residential long term more than 30 days), while 6-8 indicates outpatient care (i.e., ambulatory intensive outpatient, ambulatory non-intensive outpatient, ambulatory detoxification). Categories 1-2, representing detoxification without continuing treatment (i.e., detox 24-hour hospital outpatient, detox 24-hour free-standing residential), were excluded. After this split, we obtained two datasets with $305,450$ inpatient records and $1,048,575$ outpatient records. Feature selection, described in Section \ref{sec:featureSelection}, reduced the number of variables to $28$ used for building predictive models. For a descriptive summary, we refer the reader to Table 1 in section 4.1, and for more detailed information on the dataset and selected variables, please refer to the Appendix Table \ref{Appendix:table:descriptive_full_list}. 

\subsubsection{Target Variable: Length-of-Stay (LOS)}
\label{sec:targetVariableLOS}
We selected LOS as our target variable because it is considered to be a reliable indicator of SUD treatment outcomes in the U.S. \citep{hubbard1989treatment,zhang2003retention,proctor2014continuing}. Despite a general policy trend in health care towards reducing LOS to lessen financial burden \citep{rotter2010pathways,stone2022los}, the best outcomes are more likely to occur with longer durations of treatment \citep{bartel2020hospitals,nida2014principles}, especially in the context of SUD treatment. Typically, LOS for SUD treatment in a rehabilitation facility is longer than for inpatient hospital admissions (average of $30$ days in rehabilitation vs. $5$ days in the hospital, \cite{cdc2017health,nida2014principles}). Consulting with prior relevant literature \citep{cdc2017health,nida2014principles,mennis2016racial,baird2023determinants} and a partner SUD treatment provider, we determined optimal LOS thresholds: $>90$ days for outpatient and $>30$ days for inpatient treatment as positive outcomes. Thus, our target LOS variable is binary: (i) $>90$ days for outpatient or $>30$ days for inpatient are considered long-term stays, leading to better outcomes and encoded as $1$; (ii) $1-90$ days for outpatient or $1-30$ days for inpatient are considered short-term stays and encoded as $0$. 

\subsection{Feature Selection}
\label{sec:featureSelection}
Selecting a concise set of features reduces model complexity and lowers computational costs while improving performance and explainability \citep{li2017feature,dunn2021feature,moreno2021explainability}. For feature selection, we utilize the Least Absolute Shrinkage and Selection Operator (LASSO), decision tree, random forest, and extreme gradient boosting (XGBoost). Among the variable sets created by each of the four feature selection algorithms, we apply a majority vote approach to the variables. A variable is included if it is selected by at least three out of the four feature selection algorithms; otherwise, it is excluded \citep{simsek2020hybrid}. 

\subsection{Computational Modeling}
\label{sec:compModeling}
The datasets we use in this study are structured and tabular. Prior work shows that tree-based algorithms typically outperform others for classification tasks involving tabular data, while recent neural network-based algorithms are emerging with comparable performance \citep{grinsztajn2022tree,borisov2021deep}. In this study, we experimented with random forest, XGBoost, LightGBM, and TabNet \citep{arik2021tabnet} to predict LOS. The first three algorithms have been used in prior studies with TEDS-D and similar datasets, justifying their inclusion in our study. In addition, we included TabNet as a neural network-based algorithm designed for tabular data, given its comparable performance with tree-based algorithms.

The class distribution in the TEDS-D dataset, given our target variable formulation discussed earlier, is $33.04\%$ longer-term ($100,924$ data points) vs. $66.96\%$ shorter-term ($204,526$) for inpatients and $35.72\%$ longer-term ($374,573$) vs. $64.28\%$ shorter-term ($674,002$) for outpatients. We split the datasets into $70\%$ for training and $30\%$ testing in a stratified manner. As we observed an imbalance in the datasets, we applied resampling methods, including the Synthetic Minority Oversampling Technique (SMOTE), on the training set. Then, the inpatient training set had a balanced distribution of classes with $101,428$ data points each, and the outpatient training set had $414,875$ data points for each class. We tuned hyperparameters and trained using $10$-fold stratified cross-validation. 

\subsection{Assessing Unfairness in ML Models for SUD Patients}
\label{sec:assessFairnessML}
To test for fairness in the final predictive models selected, we utilized FairLearn\footnote{  \texttt{https://fairlearn.org/}}. FairLearn provides fairness metrics to help identify which social groups may be adversely impacted by models \citep{hardt2016equality,bird2020fairlearn}. Our process involves the use of FairLearn to: (i) assess which patient characteristics are associated with negative impacts due to model predictions and then (ii) compare multiple models in terms of their fairness. 

Following prior seminal studies on the interaction between protected (e.g., race and socioeconomic) and the target variables, we apply the criterion that the probability of a positive outcome should be the same regardless of the population group membership \citep{cleary1968item,thorndike1971culture,scheuneman1979bias,hutchinson2019test}. In other words, the ratio of predicted positive outcomes to ground truth positive outcomes should be equal for each group, which is called the \emph{selection rate}. We specifically use fairness metrics to measure model outcomes for parity in the selection rates across all subgroups in a variable. More specifically, we use selection rate, demographic parity, equalized odds, false positive rate difference, and false negative rate difference \citep{castelnovo2022fairness}. These metrics were applied when evaluating fairness for groups within three variable categories: demographics, medical, and financial. Demographic variables include race, marital status, and U.S. region. For medical variables, we consider SUD diagnosis, mental health comorbidity, medication-assisted opioid therapy, substance use at discharge, and admission wait time. For financial variables, we consider payment source, income source, employment, living arrangements at discharge, and health insurance.

Our assessment strategy is two-fold: (i) group-level (i.e., variables), and (ii) subgroup-level (i.e., categories within variables) evaluations. Group-level fairness assessment evaluates models using metrics such as demographic parity, equalized odds, false positive rate difference, false negative rate difference, and overall selection rate. These metrics range from 0 to 1, with values closer to either end of the continuum indicating higher bias against a certain group. Groups identified as potentially at risk of discrimination based on these metrics are further investigated at the subsequent subgroup-level assessment. Subgroup-level fairness assessment involves calculating selection rates and demographic parity ratios across subgroups within each variable, measuring model performance parity. The demographic parity ratio is the ratio between minimum and maximum selection rates. We also calculate this ratio for each subgroup with respect to the subgroup with the maximum selection rate. Guidelines\footnote{Disparate Impact. Griggs v. Duke Power Co., 401 U.S. 424 (1971).} suggest a demographic parity ratio above $0.80$ for fairness (e.g., Disparate Impact with $80\%$ rule, \cite{stephanopoulos2018disparate,barocas2016disparate,wax2011disparate}). More specifically, this guideline states, “… a selection rate for any race, sex, or ethnic group which is less than four-fifths... of the rate for the group with the highest rate will generally be regarded by the Federal enforcement agencies as evidence of adverse impact...” \citep{holzer2000affirmative}. Hence, we identify subgroups with ratios below this $80\%$ threshold. 

%% file: 4-Results.tex

In this section, we present the results of our approach for feature selection, modeling for classification, and fairness assessment for our models. As we outlined in Section \ref{sec:targetVariableLOS}, the positive outcome (i.e., $1$) corresponds to a longer-term LOS, while a negative outcome (i.e., $0$) represents a shorter-term LOS.

\subsection{Feature Selection Results}
\label{sec:resultsFeatureSelect}
We performed a feature selection procedure that is described in Section \ref{sec:featureSelection}. Via feature selection, we selected $28$ variables in that we then used for training the ML models. Table 1 below describes a subset of $8$ variables selected by all four algorithms in our feature selection process. We provide a descriptive summary of these variables here and refer the reader to the full list of selected variables and their descriptions in the Appendix Table \ref{Appendix:table:descriptive_full_list}. 

\begin{table}[h!]
\centering
\begin{tabular}{p{2cm}p{10cm}}
\toprule[2pt]
\textbf{Variables} & \textbf{Description} \\ 
\midrule[1pt]
METHUSE & Medication-assisted opioid therapy such as methadone, buprenorphine, and/or naltrexone \\ 
\midrule[1pt]
ARREST & Arrests in past 30 days prior to admission \\ 
\midrule[1pt]
MARSTAT & Marital status \\ 
\midrule[1pt]
LIVARAG\_D & Living arrangements at discharge; homeless, a dependent or living independent \\ 
\midrule[1pt]
PSOURCE & Referral source \\ 
\midrule[1pt]
SUB1\_D & Substance use at discharge (primary) \\ 
\midrule[1pt]
DAYWAIT & Days waiting to enter substance use treatment \\ 
\midrule[1pt]
Division & Census region \\ 
\bottomrule[2pt]
\end{tabular}
\caption{Descriptive Summary for TEDS-D. These variables are presented in this table as per their importance and relevance to the fairness evaluation. \textit{These variables were also selected by all four algorithms in our feature selection process.} The full list of selected variables and their descriptive summary is provided in the Appendix Table \ref{Appendix:table:descriptive_full_list}.}
\label{table:descriptive_summary}
\end{table}

\subsection{Evaluation of ML Models}
\label{sec:resultsMLmodels}
We evaluate the performance of the predictive models trained via random forest, LightGBM, XGBoost, and TabNet through accuracy, precision, recall, and F1 score. We report the classification results in Table \ref{table:classification_results} for the models trained over the TEDS-D dataset for both inpatient and outpatient. We observed that the random forest and LightGBM models consistently outperformed XGBoost and TabNet for both inpatient and outpatient data from the TEDS-D dataset. Notably, models for inpatient data overall performed better. We selected random forest and LightGBM, which utilize ensemble methods based on decision trees, as the best-performing two models for fairness assessment. Nevertheless, we recognize the comparable results that XGBoost and TabNet produce. 

\begin{table}[h!]
\centering
\begin{tabular}{p{2.2cm}p{0.9cm}p{0.9cm}p{0.9cm}p{0.9cm}p{0.9cm}p{0.9cm}p{0.9cm}p{0.9cm}}
\toprule[2pt]
\textbf{Model} & \multicolumn{2}{c}{\textbf{Accuracy}} & \multicolumn{2}{c}{\textbf{Precision}} & \multicolumn{2}{c}{\textbf{Recall}} & \multicolumn{2}{c}{\textbf{F-1}} \\ 
\cmidrule[1pt]{2-9}
 & \textbf{InP} & \textbf{OutP} & \textbf{InP} & \textbf{OutP} & \textbf{InP} & \textbf{OutP} & \textbf{InP} & \textbf{OutP} \\ 
\midrule[1pt]
XGBoost & 0.81 & 0.75 & 0.81 & 0.75 & 0.81 & 0.75 & 0.81 & 0.75 \\ 
\midrule[1pt]
\rowcolor[gray]{0.8} LightGBM & 0.81 & 0.76 & 0.81 & 0.76 & 0.81 & 0.76 & 0.81 & 0.76 \\ 
\midrule[1pt]
\rowcolor[gray]{0.8} Random Forest & 0.81 & 0.77 & 0.81 & 0.77 & 0.81 & 0.77 & 0.81 & 0.77 \\ 
\midrule[1pt]
TabNet & 0.79 & 0.73 & 0.78 & 0.74 & 0.78 & 0.74 & 0.78 & 0.74 \\ 
\bottomrule[2pt]
\end{tabular}
\caption{Classification results. The classifiers in gray are selected as best-performing models; hence, to be assessed for fairness. \textbf{InP}: Inpatient, \textbf{OutP}: Outpatient.}
\label{table:classification_results}
\end{table}

\subsection{Fairness in ML Models}
\label{sec:resultsFairnessML}
The results in sections \ref{sec:resultsFairnessInPatientGroupLevel} and \ref{sec:resultsFairnessOutPatientGroupLevel} provide a comprehensive view of the fairness assessment performed on the two, selected ML classification models (LightGBM, random forest) trained to predict the LOS for patients in SUD treatment programs. This assessment approach utilized several metrics to evaluate model fairness, namely, Equalized Odds Difference (EOD), False Positive Rate Difference (FPRD), False Negative Rate Difference (FNRD), Overall Selection Rate (OSR), and Demographic Parity Ratio (DPR) \citep{castelnovo2022fairness}. These metrics collectively help determine if the models exhibit any bias based on the selected variables. As we develop models for inpatient and outpatient data, we assess these models separately. 

\subsubsection{Group-level Fairness Assessment of the Inpatient Models}
\label{sec:resultsFairnessInPatientGroupLevel}
As described in section \ref{sec:assessFairnessML} and detailed in the Appendix Section \ref{Appendix:sec:fairnessMetrics}, the equalized odds metric measures the bias in the likelihood of a predicted positive outcome being uniform across all groups for each examined variable. Concurrently, the equalized odds difference is utilized to gauge the level of parity among these groups. The expectation is that for a given variable, the bias should be consistent across groups, leading to an equalized odds difference approaching zero, which indicates minimal bias. In Table \ref{table:fairness_inpatients}, we see that the LightGBM model exhibits an overall lower equalized odds difference compared to the random forest model. A nonzero equalized odds difference for a protected variable suggests a discrepancy in the model's ability to accurately predict the positive outcome equally across all groups. With the positive outcome being a longer-term stay, the models with elevated equalized odds differences are prone to generating false positives. This implies a bias towards certain groups, potentially resulting in unnecessarily prolonged treatment programs and delayed discharges, which could have negative financial and possibly medical consequences for the affected patients. Differences in false positive and negative rates indicate variations in the accuracy of predictions for positive and negative outcomes among different groups. Consistent low values, as seen in ‘Health Insurance’, ‘Income Source’, and ‘Opioid Therapy’ imply more equitable false prediction rates. On the other hand, ‘Race’ and ‘US Region’ shows higher differences, especially for LightGBM, which could mean that patients associated with subgroups associated with race and US region are more likely to be misclassified. As the demographic parity ratio represents how evenly the outcomes are distributed across groups, a value of $1$ would indicate perfectly fair parity. Ratios significantly different from $1$ (e.g., race and US region) suggest that the model's outcomes are disproportionately distributed among these groups. From the models for inpatient data, there seem to have been notable variables, such as race and US region, where the fairness metrics indicate potentially significant biases. Further, the groups of SUD diagnosis, substance use at discharge, admission wait time, marital status, living arrangement at discharge also show potential bias, albeit not as significant as race and U.S. region. These findings warrant further investigation on subgroups within these groups of patients who may disproportionately experience unfair treatment in the form of unequal treatment durations.

\begin{table}[h!]
\centering
\begin{tabular}{p{3cm}p{0.5cm}p{0.5cm}p{0.5cm}p{0.5cm}p{0.5cm}p{0.5cm}p{0.5cm}p{0.5cm}p{0.5cm}p{0.5cm}}
\toprule[2pt]
\textbf{Variables} & \multicolumn{5}{c}{\textbf{LightGBM}} & \multicolumn{5}{c}{\textbf{Random Forest}} \\ 
\cmidrule(lr){2-6} \cmidrule(lr){7-11}
\textbf{\emph{Inpatient}}  & EOD & FPRD & FNRD & OSR & DPR & EOD & FPRD & FNRD & OSR & DPR \\
\midrule[1pt]
SUD Diagnosis & 0.38 & 0.33 & 0.38 & 0.57 & 0.39 & 0.36 & 0.27 & 0.36 & 0.61 & 0.47 \\
\rowcolor[gray]{0.8} Mental-SUD Comorbid & 0.13 & 0.08 & 0.13 & 0.57 & 0.70 & 0.15 & 0.15 & 0.15 & 0.61 & 0.68 \\
Medical Assisted Opioid Therapy & 0.07 & 0.07 & 0.04 & 0.57 & 0.85 & 0.11 & 0.11 & 0.07 & 0.61 & 0.82 \\
\rowcolor[gray]{0.8} Substance Use at Discharge & 0.35 & 0.28 & 0.35 & 0.57 & 0.43 & 0.38 & 0.38 & 0.34 & 0.61 & 0.50 \\
Admission Wait Time & 0.23 & 0.16 & 0.23 & 0.57 & 0.40 & 0.26 & 0.22 & 0.26 & 0.61 & 0.39 \\
\rowcolor[gray]{0.8} Race & 0.84 & 0.46 & 0.84 & 0.57 & 0.27 & 0.96 & 0.96 & 0.94 & 0.61 & 0.19 \\
Marital Status & 0.28 & 0.16 & 0.28 & 0.57 & 0.40 & 0.27 & 0.27 & 0.27 & 0.61 & 0.45 \\
\rowcolor[gray]{0.8} Health Insurance & 0.15 & 0.15 & 0.08 & 0.57 & 0.75 & 0.12 & 0.12 & 0.07 & 0.61 & 0.78 \\
Living Arrangement at Discharge & 0.26 & 0.16 & 0.26 & 0.57 & 0.57 & 0.25 & 0.22 & 0.25 & 0.61 & 0.57 \\
\rowcolor[gray]{0.8} US Region & 0.44 & 0.40 & 0.44 & 0.57 & 0.28 & 0.58 & 0.58 & 0.47 & 0.61 & 0.19 \\
Employment & 0.21 & 0.21 & 0.14 & 0.57 & 0.58 & 0.23 & 0.23 & 0.17 & 0.61 & 0.58 \\
\rowcolor[gray]{0.8} Income Source & 0.14 & 0.14 & 0.13 & 0.57 & 0.62 & 0.14 & 0.14 & 0.11 & 0.61 & 0.61 \\
Payment Source & 0.25 & 0.25 & 0.11 & 0.57 & 0.63 & 0.25 & 0.25 & 0.13 & 0.61 & 0.62 \\
\bottomrule[2pt]
\end{tabular}
\caption{Fairness Assessment Group-level Results for TEDS-D \textit{Inpatients} based on Demographic, Medical and Financial variables. EOD: Equalized Odds Difference, FPRD: False Positive Rate Difference, FNRD: False Negative Rate Difference, OSR: Overall Selection Rate, DPR: Demographic Parity Ratio.}
\label{table:fairness_inpatients}
\end{table}

\subsubsection{Group-level Fairness Assessment of the Outpatient Models}
\label{sec:resultsFairnessOutPatientGroupLevel}
The overall selection rate represents the proportion of instances predicted as a positive outcome, which, in this study, pertains to the predicted longer stay for SUD patients. The random forest model consistently demonstrated a more equitable distribution of predicted positive outcomes compared to the LightGBM model. Given that we consider lower values in false negative rate differences and equalized odds difference to be indications of a fairer model, the random forest model exhibits values nearer to zero across these metrics, suggesting it is the more equitable model for the selected variables. As the demographic parity ratio provides insight into the distributional equality of the model's predictions (i.e., $\sim1$ signifies fair), the results in Table \ref{table:fairness_outpatients}, the random forest model approaches $1$ more closely than LightGBM, implying a better balance in fairness. Further, the race variable shows the highest FNR differences, suggesting bias against certain racial subgroups and potential impact in terms of medical implications. This situation is particularly concerning since misclassification of positive outcomes as negative can lead to premature discharges, resulting in potentially adverse health consequences for certain racial subgroups, potentially with a risk of relapse of the disorder. The analysis of these metrics for outpatient models indicates that certain variables exhibit significant levels of unfairness within both models, including race, substances used at discharge, payment source, and U.S. region with higher bias.

\begin{table}[h!]
\centering
\begin{tabular}{p{3.3cm}p{0.5cm}p{0.5cm}p{0.5cm}p{0.5cm}p{0.5cm}p{0.5cm}p{0.5cm}p{0.5cm}p{0.5cm}p{0.5cm}}
\toprule[2pt]
\textbf{Variables} & \multicolumn{5}{c}{\textbf{LightGBM}} & \multicolumn{5}{c}{\textbf{Random Forest}} \\ 
\cmidrule(lr){2-6} \cmidrule(lr){7-11}
\textbf{\emph{Outpatient}} & EOD & FPRD & FNRD & OSR & DPR & EOD & FPRD & FNRD & OSR & DPR \\
\midrule[1pt]
SUD Diagnosis & 0.26 & 0.17 & 0.26 & 0.58 & 0.54 & 0.154 & 0.154 & 0.152 & 0.65 & 0.64 \\
Mental-SUD Comorbid & 0.21 & 0.09 & 0.21 & 0.58 & 0.66 & 0.13 & 0.09 & 0.13 & 0.65 & 0.72 \\
Opioid Therapy & 0.22 & 0.17 & 0.22 & 0.58 & 0.56 & 0.18 & 0.18 & 0.11 & 0.65 & 0.65 \\
\rowcolor[gray]{0.8} Substance Use at Discharge & 0.43 & 0.43 & 0.36 & 0.58 & 0.20 & 0.48 & 0.48 & 0.32 & 0.65 & 0.22 \\
Admission Wait Time & 0.34 & 0.11 & 0.34 & 0.58 & 0.48 & 0.19 & 0.11 & 0.19 & 0.65 & 0.57 \\
\rowcolor[gray]{0.8} Race & 0.39 & 0.39 & 0.37 & 0.58 & 0.36 & 0.39 & 0.39 & 0.27 & 0.65 & 0.43 \\
Marital Status & 0.07 & 0.07 & 0.07 & 0.58 & 0.86 & 0.07 & 0.07 & 0.07 & 0.65 & 0.88 \\
Health Insurance & 0.22 & 0.17 & 0.22 & 0.58 & 0.58 & 0.20 & 0.12 & 0.12 & 0.65 & 0.64 \\
Living Arrangement at Discharge & 0.16 & 0.16 & 0.14 & 0.58 & 0.72 & 0.20 & 0.20 & 0.08 & 0.65 & 0.74 \\
\rowcolor[gray]{0.8} US Region & 0.47 & 0.32 & 0.47 & 0.58 & 0.32 & 0.29 & 0.29 & 0.28 & 0.65 & 0.44 \\
Employment & 0.22 & 0.11 & 0.22 & 0.58 & 0.62 & 0.115 & 0.111 & 0.115 & 0.65 & 0.71 \\
Income Source & 0.15 & 0.12 & 0.15 & 0.58 & 0.73 & 0.09 & 0.09 & 0.07 & 0.65 & 0.81 \\
\rowcolor[gray]{0.8} Payment Source & 0.51 & 0.30 & 0.51 & 0.58 & 0.30 & 0.37 & 0.37 & 0.25 & 0.65 & 0.44 \\
\bottomrule[2pt]
\end{tabular}
\caption{Fairness Assessment Group-level Results for TEDS-D \textit{Outpatients} based on Demographic, Medical, and Financial variables. EOD: Equalized Odds Difference, FPRD: False Positive Rate Difference, FNRD: False Negative Rate Difference, OSR: Overall Selection Rate, DPR: Demographic Parity Ratio.}
\label{table:fairness_outpatients}
\end{table}

We further note that equalized odds difference values in the outpatient models tend to be lower than those in the inpatient models. This indicates that, for outpatient predictions, both the LightGBM and random forest models may be achieving a more uniform performance across groups, potentially due to the nature of the outpatient data or the variables’ impact in that context. False positive and false negative rate differences for inpatient models show higher disparities across several variables, suggesting that inaccuracies in the prediction of the length of stay are more pronounced in the inpatient context. This might be due to the complexity of inpatient cases or the greater variability in the types of treatments and interventions administered. Outpatient models generally exhibit higher demographic parity ratio values closer to $1$, which suggests that the predictions are more evenly distributed among different groups compared to inpatient models. These demographic parity ratios in outpatient settings might reflect less complex social dynamics or a more consistent application of treatment protocols. The reasons behind these comparisons may lie in the inherent differences between inpatient and outpatient care. Inpatient treatment often involves more acute, severe, and varied cases, which could introduce greater unpredictability into model predictions and thus affect fairness metrics. On the other hand, outpatient care might involve more standardized and less variable treatment pathways, facilitating more uniform predictions across groups. These insights indicate that while both inpatient and outpatient models are designed to serve similar predictive purposes, the context in which they operate significantly impacts their fairness. Inpatient models may require more nuanced training and greater attention to complex variables to enhance fairness, while outpatient models might benefit from refining existing variables to ensure consistent fairness across groups.

\begin{figure}[h!]
    \centering
    \includegraphics[width=12cm]{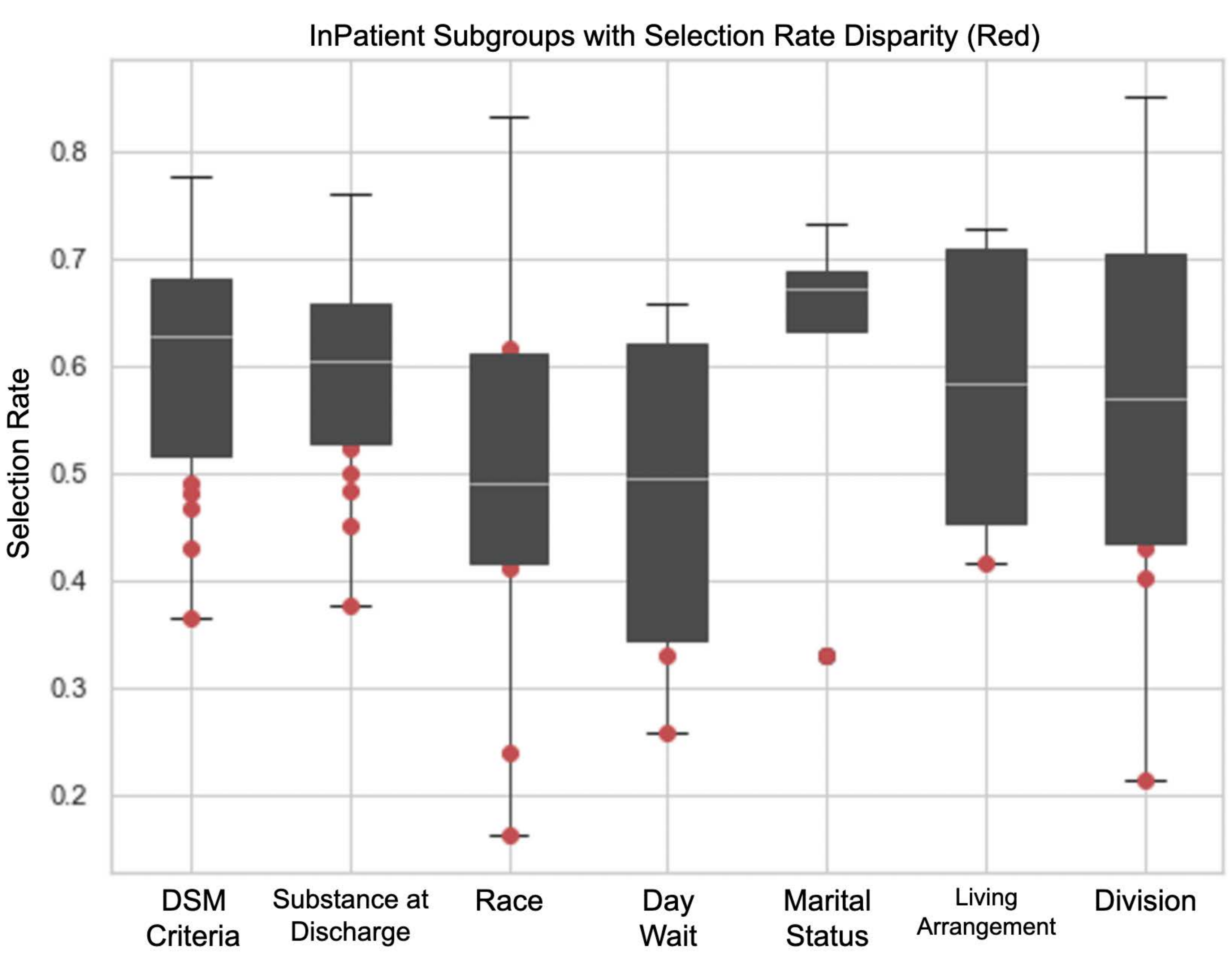}  
    \caption{Boxplot of inpatient selection rate parity across various variables highlighting subgroups with significantly lower rates (red dots).}
    \label{fig:inpatientDotplot}
\end{figure}

\subsubsection{Subgroup-level Fairness Assessment of the Inpatient Models}
\label{sec:resultsFairnessSubgroupInpatient}
By calculating the selection rates for each subgroup, our goal was to identify those most vulnerable to potential unfair treatment by the predictive models. Fairness in these models is gauged by the equality of selection rates; hence, subgroups with significantly lower rates are signs of bias. Figure \ref{fig:inpatientDotplot} presents a box plot where there is a range of selection rates for each variable, indicated by the height of the boxes. The central line in each box represents the median selection rate, while the edges of the box indicate the interquartile range (IQR), spanning the middle $50\%$ of the data. The whiskers extend to the rest of the distribution except for points that are determined to be outliers, disadvantaged subgroups. 

The red dots, which signify subgroups with highly lower selection rates compared to others within the same variable, determined by the threshold (i.e., $0.80$) based on the demographic parity ratio, indicate potential areas of disparity. It is crucial to identify these subgroups as they may be underserved by the model. Variables, including SUD diagnosis, US region, race, and substance use at discharge, show a broader range of selection rates, suggesting a higher variability in how different subgroups within these variables are selected by the model. The admission wait-times and marital status variables show less variation in selection rates, but still include subgroups with much lower selection rates, implying those subgroups may not be equitably predicted by the model. The presence of subgroups with lower selection rates raises concerns about the fairness of the model, as some of these subgroups are consistently less likely to be selected for longer stays.

More specifically, our analysis, as illustrated in the results (see Figure \ref{fig:inpatientDotplot} and Appendix Tables \ref{Appendix:table:inpatient_sr_values-6}-\ref{Appendix:table:living_arrangement_discharge-11}), reveals particular concern among inpatient subgroups with SUD diagnoses, specifically those with "Cannabis Abuse" and "Attention Deficit/Disruptive Behavior Disorders," along with a subgroup with missing or unknown diagnoses, all of which exhibit lower selection rates. Notably, the subgroup with "Attention Deficit/Disruptive Behavior Disorders" is also underrepresented in the dataset, which suggests a broader issue of representation. Such underrepresentation presents a challenge not only in societal terms but also from a technical perspective. Appendix Table \ref{Appendix:table:substance_discharge_inpatient-7} shows that subgroups of patients with marijuana and other drug use are associated with notably lower selection rates (i.e., red dots in Figure \ref{fig:inpatientDotplot}). This suggests that such groups are disproportionately affected and may not be adequately served by the models. For example, predicted shorter than necessary stays for treatment (representing false negatives) imply that patients with addiction to marijuana or other drugs could be prematurely discharged, potentially depriving them of the necessary duration of care for effective recovery.

Recall that Table \ref{table:fairness_inpatients} shows that the race variable stands out with the highest disparities in fairness metrics for inpatient groups, revealing more biases affecting certain racial subgroups. Further analysis presented in Appendix Table \ref{Appendix:table:race_inpatient-8} highlights that underrepresented subgroups tend to have relatively lower selection rates. These substantial discrepancies in selection rates contribute to the model's bias against these subgroups. The gap in selection rates, with the highest being $0.83$ for the Asian/Pacific Islander subgroup and the lowest at $0.17$ for the Alaska Native subgroup, indicates inequity, suggesting that the model's performance is not uniformly effective across all racial subgroups. Notably, for the inpatient, the three subgroups delineated in Appendix Table \ref{Appendix:table:race_inpatient-8} and marked in red in Figure \ref{fig:inpatientDotplot}, are characterized by highly reduced selection rates, signaling a heightened risk of bias. For example, patients belonging to the Native Hawaiian Other Pacific Islander subgroup could be subject to premature discharge, resulting in potentially insufficient treatment duration for mental health care, which represents a concerning rate of false negatives.

As shown in Appendix Table \ref{Appendix:table:wait_time_admission-9}, which focuses on the wait times in days for treatment, there is a notable variation in selection rates among different subgroups, ranging from $0.26$ to $0.66$. This range signals a considerable variance in the level of bias affecting these subgroups. Notably, patients who experience wait times exceeding $8$ days are subject to the lowest selection rates. Additionally, it is observed that these subgroups with extended wait times also tend to be underrepresented, which may correlate with the observed increased bias against them.

Marital status emerges as a critical factor in predicting the length of inpatient treatment stays. Appendix Table \ref{Appendix:table:marital_status_inpatient-10} shows that most of the subgroups have seemingly similar performance by the model, while the patients with no recorded marital status information may be exposed to inequitable treatment outcomes from the predictive models, as per the lowest selection rates. Notably, while the subgroup comprising never-married patients represents the largest demographic, it is the married patients who are most frequently selected for longer stays, suggesting a potential, albeit not severe, disparity in model predictions based on marital status.

The type of living arrangement patients will return to after discharge is a crucial consideration because the inpatient treatment environment plays a significant role in the overall effectiveness of SUD treatments. According to the results in Appendix Table \ref{Appendix:table:living_arrangement_discharge-11}, patients who are homeless or living independently are more often predicted to have extended stays within treatment facilities compared to those in dependent living situations or those whose living arrangements after discharge are unspecified. Essentially, the model tends to favor longer treatment durations for patients who lack a supportive home environment or peer support. 

The U.S. region also emerges as another influential factor. Disparities in selection rates across U.S. regions are evident from the data in Appendix Table \ref{Appendix:table:us_region_inpatient-12}, as only three regions, namely East North Central, Middle Atlantic, and East South-Central regions, show the most satisfactorily equitable treatment. These findings highlight the necessity to consider both individual and systemic factors when assessing and improving the fairness of predictive models in mental health care.

\subsubsection{Subgroup-level Fairness Assessment of the Outpatient Models}
\label{sec:resultsFairnessSubgroupOutpatient}
Outpatient treatment programs for SUD have distinct attributes related to diagnosis, substance use patterns, and progression of the disorder, among other factors. Outpatient care typically caters to patients whose conditions are less severe and do not necessitate the intensive supervision provided in an inpatient setting or for those transitioning from inpatient care after achieving certain treatment milestones \citep{samhsa2024equity}. Compared to inpatient programs, outpatient settings often offer more flexibility and are designed to integrate treatment into the patient's daily life. While inpatient programs are more immersive and structured, outpatient treatment requires patients to have a stable living situation and a supportive environment as they continue to navigate day-to-day responsibilities. The models we have developed for outpatient treatment take these different dynamics into account and aim to reflect the nuanced needs of these patients. As we developed models to predict treatment length for outpatient services, we observed that variables for race, U.S. region, substance use at discharge, and payment source exhibited relatively higher bias. Hence, we focus our analysis on the subgroups within these variables to assess fairness.

\begin{figure}[h!]
    \centering
    \includegraphics[width=7.5cm]{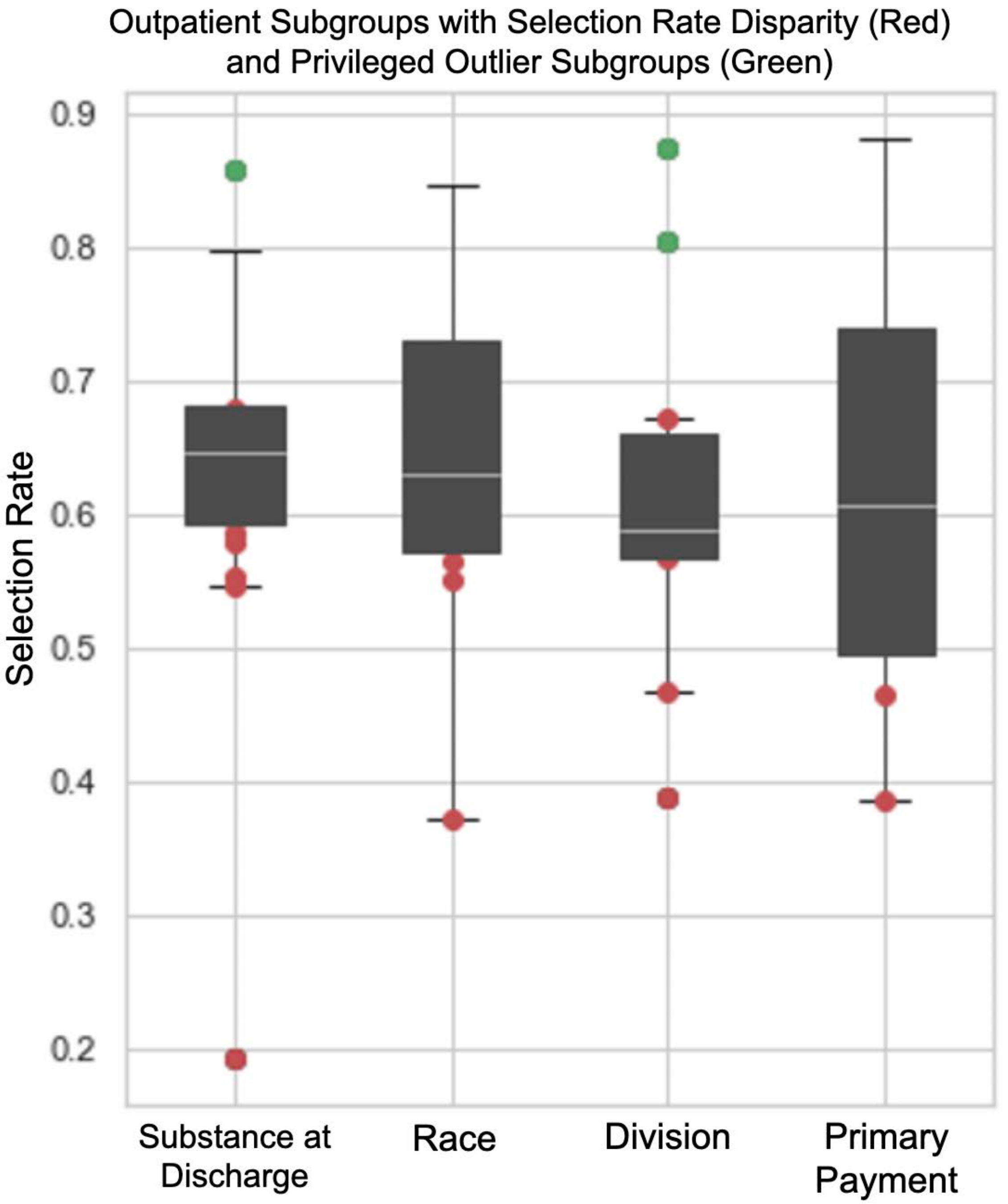}  
    \caption{Boxplot of Outpatient selection rate parity ratio across various variables highlighting subgroups with significantly lower rates (red dots) and privileged outlier subgroups (green dots).}
    \label{fig:outpatientDotplot}
\end{figure}

Figure \ref{fig:outpatientDotplot} illustrates the distribution of selection rates for outpatient subgroups across four variables: SUD diagnosis, race, U.S. region, and primary payment source. Significant disparities in selection rates across these subgroups indicates potential bias or unfairness in the model. As the red dots in the figure indicate subgroups that are selected at rates lower than the threshold ($0.80$) based on the demographic parity ratio, these subgroups may not be receiving equitable consideration by the models. They may be potentially underrepresented or underserved in the predicted treatment durations. On the other hand, we see green dots in the figure representing outliers in selection rate parity, indicating privileged subgroups that receive highly favorable predictive positive outcomes from the models. These subgroups may receive longer treatment durations, which could be beneficial for their recovery, while raising questions about fair resource allocation as to whether this reflects an actual need or a bias in the model. The substance use at discharge variable shows a relatively balanced distribution in selection rates across subgroups, but the presence of subgroups at both the lower and higher suggests the presence of disparities in treatment durations. Appendix Table \ref{Appendix:table:substance_discharge_outpatient-13} shows that patients who are using barbiturates at the time of discharge are at a heightened risk of misclassification by the model. Only a limited number of subgroups, specifically those associated with substances such as cocaine/crack, methamphetamine, other stimulants, drugs, or unknown substances, appear to be more favorably selected by the model for longer treatment stays. 

The selection rates across racial subgroups display variability (see Appendix Table \ref{Appendix:table:race_outpatient-14}), which may indicate disparities among racial subgroups based on predictive performance by the models. For instance, the Alaska Native group shows the lowest selection rate, while the subgroup with no racial information has the highest selection rate. Subgroups such as Alaska Native and Asian, among others, are at the lower end of the selection rates, which could imply a higher risk of misclassification for these subgroups. This risk could manifest as premature discharge or insufficient treatment duration, as predicted by the models. We have observed significant variability in selection rates across different racial subgroups (see Figure \ref{fig:outpatientDotplot} and Appendix Table \ref{Appendix:table:race_outpatient-14}), resulting in most subgroups, except for the top three, having selection rate parity ratios that fall below the defined threshold. This indicates that there is a disparity in the predicted LOS for these subgroups, receiving shorter stays than equitable or necessary for effective treatment. 

U.S. Regions appear to have a wider spread of selection rates with notable variability across the regions as the East South Central and Mountain regions have the highest selection rates ($>0.80$) being outliers, while the West South-Central region shows the lowest selection rates ($0.39$) (see Appendix Table \ref{Appendix:table:us_region_outpatient-15}). This disparity between U.S. regions, especially the extreme values for the East South Central and West South-Central regions, raises questions about the potential influence of regional health care practices, resource allocation, or other systemic factors on health care outcomes, including the treatment length. 

Appendix Table \ref{Appendix:table:payment_source_outpatient-16} shows the subgroups with different payment sources where the selection rates vary significantly. Patients covered by a program such as Medicare, Medicaid, or private insurance have higher selection rates, indicating bias for longer predicted stays. In contrast, the patients who pay out-of-pocket or do not pay (e.g., ‘No charge' subgroup) have the lowest selection rates, suggesting shorter predicted LOS in treatment. The disparity in selection rates based on payment source raises questions about potential financial biases in the length of treatment stays, raising concerns about equitable access to care and potentially perpetuating existing socioeconomic disparities. In other words, the lower selection rates for ‘Self-pay' and ‘No charge' subgroups highlight a risk of ML models under-predicting the needed length for treatment and care among patients who may be financially vulnerable or less able to pay. The higher selection rates for ‘Medicare' and ‘Medicaid' subgroups might reflect systemic tendencies to approve longer treatment stays for patients with established private or government insurance, which could also relate to regulatory or reimbursement factors.

\subsubsection{Summary of Results}
\label{sec:resultsSummary}
Table \ref{table:resultsSummary} summarizes the results for both SUD inpatients and outpatients, highlighting potentially disadvantaged groups and subgroups. For inpatients, significant disparities were observed across demographic, regional, and treatment-related variables. Specifically, individuals identifying as Alaska Native, Native Hawaiian, or Other Pacific Islander were disproportionately impacted, as were those from certain U.S. regions, including territories and less urbanized areas. Patients using marijuana or other drugs, those with diagnoses, such as cannabis abuse or attention-deficit/disruptive behavior disorders, and those with missing or unknown marital statuses also faced unequal treatment recommendations. Similarly, for outpatients, Alaska Native individuals and those residing in a broad range of geographic regions—including territories and rural states—emerged as potentially disadvantaged subgroups. Additionally, patients with barbiturate use at discharge and those relying on self-pay or other non-insurance payment sources were identified as at-risk groups.

\begin{table}[h!]
\centering
\begin{tabular}{p{1.3cm}p{3.6cm}p{6.9cm}}
\toprule[2pt]
\textbf{Type} & \textbf{Group-level} & \textbf{Subgroup-level} \\
\midrule[1pt]
\multirow{6}{*}{\textbf{Inpatient}} 
& Race & Alaska Native, Native Hawaiian, Other Pacific Islander \\
\cmidrule(lr){2-3}
& U.S. Region & U.S. territories, New England, West North Central, South Atlantic, West South Central, Mountain, Pacific \\
\cmidrule(lr){2-3}
& Substance Use at Discharge & Marijuana, Other drug use \\
\cmidrule(lr){2-3}
& SUD Diagnosis & Cannabis Abuse, Attention Deficit/Disruptive Behavior Disorders, Missing or Unknown diagnoses \\
\cmidrule(lr){2-3}
& Admission Wait Times & 8 days or more \\
\cmidrule(lr){2-3}
& Marital Status & Missing Unknown marital status \\
\midrule[1pt]
\multirow{4}{*}{\textbf{Outpatient}} 
& Race & Alaska Native \\
\cmidrule(lr){2-3}
& US Region & U.S. territories, New England, Middle Atlantic, East North Central, West North Central, South Atlantic, West South Central, Pacific \\
\cmidrule(lr){2-3}
& Substance Use at Discharge & Barbiturates \\
\cmidrule(lr){2-3}
& Payment Source & No charge, Other, Self-pay \\
\bottomrule[2pt]
\end{tabular}
\caption{Summary table for potentially disadvantaged groups and subgroups.}
\label{table:resultsSummary}
\end{table}

%% file: 5-Discussion.tex

In this study, we have developed and evaluated computational innovations in the form of various ML classification models, specifically to predict the LOS for patients in SUD treatment programs. Our variables span demographic, medical, and financial categories. Through the lenses of distributive justice and socio-relational fairness, we scrutinized the models for fairness at both the group and subgroup levels by utilizing fairness metrics including equalized odds, demographic parity ratios, as well as differences in false positive and false negative rates, in addition to selection rates. Our findings have shown variability in selection rates across subgroups within each variable category. Such disparities raise concerns about the potential for systemic biases reflected in predictive models, which may exacerbate existing societal inequities. Certain subgroups, particularly those with specific racial backgrounds, without insurance, with specific DSM diagnoses, or being treated in specific geographic regions, are under-selected by these models, thereby potentially denying them adequate care by providing shorter treatment stays than necessary. This divergence from optimal treatment has far-reaching implications for the well-being of these individuals and for how policy should be developed to address such biases. We first discuss how identified fairness issues may be mitigated, given the findings in this study, and then discuss implications for policymakers and practitioners.  

\subsection{Bias Mitigation for SUD Patients}
\label{sec:discussMitigation}
We now outline a mitigation approach with two levels of strategies: (i) Model Adjustment and (ii) Social Inclusion.

\subsubsection{Model Adjustment}
\label{sec:discussMitigateModelAdjust}
The technical model mitigation process involves adjusting model parameters for equity, integrating additional financial or social variables, and ensuring balanced representation in datasets. More specifically, technical model mitigation strategies can be categorized into preprocessing, in-processing, and postprocessing techniques, each with specific applications to SUD treatment predictions. Preprocessing techniques modify the training data before model development to reduce bias. This could involve reweighting or resampling data to ensure better representation of minority groups, or using data augmentation techniques to address underrepresented subgroups. For instance, incorporating social and financial variables, such as access to health care resources or employment status, can improve model fairness by providing a more comprehensive understanding of patient contexts \citep{kamiran2012classification,kursuncu2019knowledge}. Preprocessing can also involve transforming features to remove correlations between protected and non-protected attributes. In-processing methods intervene during the training to incorporate fairness constraints directly into the model optimization process. One effective in-processing approach for SUD treatment predictions could be adversarial debiasing, where a neural network classifier is trained to minimize prediction error while an adversarial network attempts to infer protected attributes, such as race or gender. The model learns to make accurate predictions without allowing the adversarial network to succeed, thereby reducing bias \citep{zhang2018adversarial}. Fairness-aware regularization can also be incorporated into the loss function to penalize unfair outcomes. Postprocessing techniques adjust model outputs to improve fairness without retraining the model. For SUD treatment models, threshold optimization can be used to align predictions with fairness criteria, such as demographic parity or equalized odds. This method ensures that the likelihood of predicting a longer or shorter length of stay is more equitably distributed among different demographic groups \citep{hardt2016equality}. These adjustments are particularly useful for ensuring that no subgroup is disproportionately burdened or privileged by the model’s recommendations.

Further, continuous monitoring and evaluation are critical to maintaining fairness throughout the model’s lifecycle. This involves tracking key fairness metrics to identify and rectify emerging biases over time \citep{barocas2016disparate}. Model auditing should be conducted periodically to ensure sustained performance and fairness, especially as new data becomes available. Automated monitoring systems can be implemented to flag significant deviations from fairness benchmarks. Additionally, community engagement and stakeholder participation are essential for the ethical assessment of these models, even once technical mitigation has been implemented. Policies for ethical oversight should be established, involving ethicists and patient representatives to review data descriptions, model design and implementation, ensuring alignment with ethical standards and public health goals.

\subsubsection{Social Inclusion}
\label{sec:discussMitigateSocialInclusion}
Improving fairness in ML models for SUD treatment goes beyond technical adjustments. It also requires a holistic approach that incorporates social inclusion and thoughtful policy development. Engaging diverse stakeholders, including individuals from marginalized communities, health care providers, data scientists, ethicists, and policymakers, is critical for designing and implementing equitable and effective ML models. This engagement helps ensure that the models reflect the needs, values, and experiences of all patient groups, particularly those who are most vulnerable or have historically faced health care disparities \citep{rajkomar2018fairness}. Representatives from marginalized groups can provide insights into potential biases and suggest culturally sensitive considerations that might otherwise be overlooked. For example, collaboration with community health organizations can reveal barriers to care that might influence model predictions, such as transportation challenges or limited access to mental health resources. This inclusion provides transparency and accountability in the model development cycles. Further, effective collaboration among data scientists, health care professionals, ethicists, and policymakers is essential to ensure that model predictions are not only accurate, but also ethically and clinically appropriate. These collaborative teams can develop guidelines for ethical AI use and help identify social determinants of health that should be incorporated into the models. For example, including data on income level, housing stability, and social support networks can lead to more equitable SUD treatment predictions \citep{irvin2020machine,he2023realworld}. Oversight committees with ethicists, clinicians, patient advocates, and data scientists can provide a structured approach to ethical governance. These committees can review model design, validate the inclusion of relevant social determinants, and monitor model performance post-deployment. They can also ensure that models are tested for biases across different patient demographics and assess whether technical mitigation strategies sufficiently address bias. Importantly, these committees can advocate for policies that prioritize health equity, such as guidelines for how ML models should be used in clinical settings and recommendations for continuous monitoring and improvement \citep{nordling2019fairer}.

\subsection{Implications for Policy}
\label{sec:discussImplicationPolicy}
Regarding demographic disparities, our analyses indicate that the subgroups identified by Asian/Pacific Islander, Native Americans, and multiple racial backgrounds show the lowest selection rates. This suggests that race and ethnicity data is especially important to collect and to account for in predictive models. Further, training models on fully representational data is especially important. Thus, policy could be developed to require that such data be collected for all admissions as well as accounted for when making LOS decisions, particularly when ML models are being used to aid such decisions. Our findings also highlight a notable variance in the US region variable, with the Pacific, New England, and Mountain regions recording the lowest selection rates, for both inpatient and outpatient treatment settings. Such variability is indicative of regional biases, suggesting that patients from these areas may be systematically underserved by the predictive models. Again, predictive models could be fairer if such data was collected and accounted for, and/or if predictive models applied in these regions were tuned on local data.

The medical disparities we identify are of particular concern, especially when considering the impact of false negatives. In our assessment, a false negative occurs when the model inaccurately predicts a patient will have a shorter stay—less than $30$ days for inpatient and less than $90$ days for outpatient—despite the patient's need for longer care. Such misclassifications by the model could result in patients being prematurely discharged, thus denying them the needed medical care. This could have serious repercussions, potentially exacerbating their condition due to insufficient treatment. Consequently, while higher false positive rates can also lead to inappropriate extensions of treatment, ultimately, it is the false negative rate differences that would shed light on the areas with unequal biases among different groups. This finding warrants further investigation into the clinical appropriateness of the predictive models and may suggest that such models in this context should be assessed according to the prevalence of false negative rates. Further, a policy could be developed to either require calibration of models according to false negative rates or could be applied by requiring statistics such as the false negative rate associated with specific patient characteristics to be displayed, thus allowing clinicians to more completely understand when predictions may be biased.

In regard to financial disparities, our study has identified that the source of payment for outpatient services and the nature of living arrangements for inpatients are variables where biases notably skew the prediction of treatment lengths. In regard to payment, disparities are most notable for patients without insurance or those paying out-of-pocket as opposed to those with Medicaid, Medicare, or private insurance. Patients covered by private insurance, Medicare, or Medicaid, appear to be favored by the models, leaving the patients with no such support or source of payment at a disadvantage. This situation is particularly concerning within the US health care system, which is predominantly privatized and requires patients to possess insurance or sufficient financial resources for adequate medical care. The ethical debate around whether insurance, financial means, or the lack thereof should influence the quality of health care one receives is an ongoing discussion surrounded by many complexities. Along this line, one area for future study would be if readmissions to SUD treatment are more likely for those with shorter than optimal LOS, which would suggest that financial burdens to society (or donors) could be lessened if LOS was considered with respect to decreasing the need for additional treatment. 

In regard to living situations, subgroups who either have dependent living situations or lack a stable living environment are adversely affected by our models' predictions. This suggests that predictive ML models in this domain might need to predict not just LOS, but also whether inpatient vs. outpatient services are more appropriate.

Finally, while a number of policy suggestions are made here, the most important policy, whether enacted as regulation or applied by SUD treatment centers in their own policies, is that bias in ML models needs to be not only identified, but also mitigated to the extent possible, as discussed in section \ref{sec:discussMitigation}. Ensuring equitable treatment for all patients, irrespective of their economic status or living conditions, is not merely a technical challenge but a moral imperative. Thus, to support equitable health care practices, policymakers must be informed by the insights gained from these oversight processes. Policy initiatives could include mandating regular audits of ML models for fairness, requiring health care organizations to disclose the use of any form of AI in treatment decisions \citep{singhal2024fairness}. Additionally, policies that promote data sharing between institutions, while respecting patient privacy, can improve the representativeness of training datasets and reduce biases in model predictions. Policymakers can also incentivize the adoption of fairness-aware algorithms and create frameworks for evaluating the societal impact of AI in health care \citep{ueda2024fairness,veale2017fairer}.

\subsection{Implications for Practice}
\label{sec:discussImplicationPractice}
For health care practitioners, these insights emphasize the need to assess the ML models with a critical eye before their deployment, utilizing an ongoing evaluation approach for fairness and accuracy. These models must be inherently designed to counteract, rather than perpetuate, the disparities that exist within our health care systems and society at large. For medical professionals, these insights necessitate a critical approach to the use of predictive models, ensuring that treatment decisions are informed by an awareness of potential biases. The nuanced interpretations of our results underscore the critical need for greater transparency, fairness, and accountability in the deployment of machine learning models within the mental health care sector. The potential for AI to offer highly desirable benefits is significant; however, it carries the risk of reinforcing societal disparities if not carefully monitored and regulated. To mitigate these risks, it's imperative to enhance data collection processes to ensure diverse and representative datasets while designing fairness-aware algorithms that account for biases detected in data and models. Further, achieving the goal of fairness for predictive models demands the formation of multidisciplinary oversight committees. These committees should bring together experts from various related fields, including data science, clinical medicine, ethics, and patient advocates to review, evaluate and oversee the use of predictive models in health care settings. 

Further, operations and resource allocations within treatment facilities may be significantly influenced by these models. Hence, awareness and continuous training on the inherent limitations and potential biases of algorithmic decision-making tools, are crucial for health care providers as well as facility management. Health care providers might need to reevaluate how they integrate predictive analytics into their workflows and day-to-day operations to ensure that their reliance on such tools does not perpetuate unfair practices. At a broader societal level, the complex implications of these predictive models create a pressing need for policy-makers to establish ethical standards and a comprehensive regulatory framework that guides the development and application of AI in health care, ensuring that it serves the interests of all patient subgroups equitably. Further, to improve the fairness of both inpatient and outpatient predictive models, we must consider the specific challenges inherent to each setting. Strategies might include incorporating more diverse data, addressing potential biases in variable selection, and considering the unique aspects of patient care in each setting.

\subsection{Implications Beyond SUD Treatment}
\label{sec:discussImplicationBeyond}
This study can be applied and generalized by researchers and practitioners in several ways, extending the applications beyond our specific focus on treatment lengths for SUD patients, such as in emerging science \citep{kuhlmann2019tentative} and innovations \citep{devasconcelos2025failure}. Concerning academic research, the approaches applied here can also be adapted for broader medical decision-making scenarios, such as allocating medical resources efficiently or managing patient cases more equitably. In general, as many decision-making models may incur bias in firms’ management decisions and related operations, similar methods can be used to derive fairer predictions, which can be robust and applicable to a large body of areas. Concerning innovations in practice, these methods can be used for enhanced fairness in ML-based feedback in clinical decision support systems used by clinicians. By incorporating fairness assessments into routine model development and ongoing refinement processes, data scientists in health care can advance the objectivity and equity of clinical decision-making, toward the goal of improving fairness in clinical decision-making. Health care providers, on the other hand, can foster a more equitable health care environment by integrating regular fairness assessments into their decision-making routines—using methods akin to those we have proposed. This commitment to fairness can ultimately lead to enhanced patient care and contribute positively to societal welfare. Such conscientious practices in model design and application can serve as a catalyst for fairness and transparency, advancing our society where the technology is guided for equitable access.

\subsection{Limitations and Future Research}
\label{sec:discussImplicationLimitationFuture}
This study includes limitations primarily related to the nature of the data utilized. The TEDS-D data offers a comprehensive geographic coverage (i.e., the majority of the U.S.), while the data is characterized by its categorical nature and the limited granularity of information, a result of necessary de-identification and standardization requirements. These conditions pose challenges to capturing the full complexity of substance use disorder treatments and patient needs. Hence, future work could expand upon our findings by incorporating more in-depth datasets, potentially directly from health care and treatment facilities. Such data could provide richer insights into patient demographics, treatment nuances, and outcomes, thereby enabling a more nuanced analysis of fairness in predictive modeling. Additionally, future research should explore the practical implementation of fair prediction models within clinical settings. It would be particularly valuable to assess the tangible impacts of utilizing unbiased predictive insights on both health outcomes and financial burdens for patients. This entails not only developing de-biased models but also closely examining their adoption by clinicians and other decision-makers involved in patient discharge processes. Moreover, an intriguing avenue for further investigation involves considering fairness in relation to projected disease progression \citep{mccradden2020ethical}. Understanding how a patient's clinical condition might influence LOS decisions adds another layer of complexity to the fairness discourse. This approach recognizes that equitable treatment extends beyond demographic parity to include the nuances of individual medical circumstances. Finally, while this study lays foundational work in assessing fairness in machine learning models for substance use disorder treatment predictions, it also highlights the critical need for ongoing research. Future work can significantly advance our understanding and implementation of equitable health care practices by embracing the practical application of fair models and acknowledging the clinical context of treatment decisions.

%% file: 6-Conclusion.tex
In conclusion, fairly determining how long patients should stay for treatment is challenging. In particular, demographic, medical, and financial considerations must be balanced in a way that does not treat patients with similar needs in biased ways. This study takes an important step toward identifying sources of bias in SUD LOS prediction models. We hope that the approaches proposed here can assist future researchers and data scientists in developing predictive algorithms, medical practitioners in making medical decisions, and policymakers in recognizing and better mitigating input bias. Overall, these findings expand our knowledge of how to responsibly manage innovations \citep{pandza2013strategic,owen2021organisational}, how to equitably deploy such innovations \citep{uddin2024fairness}, and contribute to our understanding of the governance of AI/ML, especially when ethical issues are present \citep{goos2024governance,reddy2020governance}. By shedding light on potentially disadvantaged subgroups, we contribute to the current discourse and body of research on health equity and social justice \citep{rajkomar2018fairness}. 

%% file: 7-Appendix.tex
\begin{appendices}

\section{Feature Selection Algorithms}
\label{Appendix:sec:featureSelection}
Wrapper methods perform search on the variable (feature) space considering all possible combinations of variables each of which is evaluated for their performance in a classification algorithm. These methods are used along with a ML algorithm in an iterative process producing models with different sets of variables. We use genetic algorithm \citep{yang1998feature} as a member of wrapper-based feature selection techniques. Filter methods utilize univariate statistical techniques to extract the intrinsic characteristics of the variables \citep{saeys2007review} based on their statistical relationship with the target variable. We use Chi-square Test \citep{li2017feature} as a member of filter-based methods. The embedded techniques are built-in methods that integrate the feature selection into the training process of the model, such as Gini index (e.g., tree-based) and regularization (e.g., LASSO) that we utilize in this study. In the following, we provide details on these feature selection methods. 

The genetics algorithm is a technique inspired by biological evolution processes to select features optimizing a given function (i.e., objective function) that represents fitness \citep{goldberg1989genetic}. In the biological analogue, a population is composed of individuals, and the fittest individuals are selected to reproduce; while in feature selection for ML, a set of solutions (population) comprises of individuals each of which represents a solution (e.g., a model) and the individuals with better fitness values based on the fitness function are selected to go through a cross-over process. Each individual is represented with a list of variables, which are swapped through crossover and changed through mutation processes. As multiple generations are created through iteratively running this process, the best-performing individual model converges with the list of variables to be selected. 

Chi-square is another statistical feature selection technique whereby a Chi-square score is computed through the assessment of whether a variable is independent of the target variable (i.e., the test of independence). This score is computed for each variable indicating the importance of variables with an alpha $0.05$ \citep{li2017feature,liu1995chi2}. The importance of a variable is determined based on its dependence on the target variable.

On the other hand, Lasso is a regularization algorithm that applies a penalty to the weights of a model to avoid overfitting, while the weights for some of the variables may be forced to go toward zero or become exactly zero \citep{li2017feature}. This property allows researchers to use Lasso to select variables with higher weights signifying the importance of the variable \citep{tibshirani1996lasso,zhu2003svm}.

Tree-based strategies for feature selection rank nodes based on the Gini index (impurity) metric that measures the impurity in data \citep{gini1971variability,li2017feature}. More specifically, the Gini index calculates the overall probability of a variable being misclassified when selected randomly. Some variables with all values the same would be considered pure, otherwise impure; hence, the Gini index indicates the degree of impurity based on the distribution of values in a variable. The lower Gini index value indicates more relevant and important features; hence, within a decision tree model, the trees below the nodes with a lower Gini index are pruned, creating a set of variables with more importance.

\section{Classification Algorithms}
\label{Appendix:sec:classificationAlgo}
Random Forest is an ensemble learning method that combines multiple models (e.g., decision trees) in parallel through bagging to select the most popular result. More specifically, multiple models are created based on a randomly chosen portion of the dataset, and among the outcome of the models generated. Since the sampling with replacement process is utilized, each model is trained over different subsets of the dataset, producing different results. a majority vote is taken to determine the final outcome for classification. XGBoost, also known as the eXtreme Gradient Boosting algorithm, is also an ensemble method that utilizes gradient boosting of decision trees combining multiple models sequentially with an additive technique \citep{chen2016xgboost}. More specifically, residual errors for the models are produced and used for training the new model while minimizing the loss using gradient descent. In recent years, XGBoost has been widely recognized for its performance over others including the neural networks on structured, tabular datasets. LightGBM (Light Gradient Boosting Machine) is another ensemble method that uses gradient boosting of decision trees, similar to XGBoost. It builds multiple models sequentially with an additive technique, where each new model corrects the errors of the previous ones \citep{ke2017lightgbm}. LightGBM employs a unique leaf-wise tree growth strategy, growing trees by splitting the leaf with the maximum delta loss, which enhances its ability to capture complex patterns. It uses histogram-based algorithms for efficient computation, making it faster and more memory-efficient than other gradient boosting methods. Recently, LightGBM has gained recognition for its superior performance on large, structured, tabular datasets, often outperforming traditional gradient boosting and neural network models. TabNet is a recent canonical deep neural network architecture for structured, tabular data using a sequential attention mechanism to select variables at each decision step allowing the identification of more important variables \citep{arik2021tabnet}. Its architecture consists of an encoder and decoder where feature transformers with multiple layers process the features which are then split and sent to the attentive transformer and the overall output. The output of the attentive transformer is then sent to a masking mechanism that provides interpretable information about the features. The masks are aggregated to learn the global importance of the features. As each such block represents a decision step and is integrated with each other, the architecture can provide an output manifold similar to the decision trees, which provides comparable overall performance with tree-based models.

\section{Fairness metrics}
\label{Appendix:sec:fairnessMetrics}
\textbf{Selection rate} is the fraction of data in each group of a variable predicted as positive outcome (i.e., $1$) in binary classification. Given $Y$ is actual target variable, $\hat{Y}$ is the predicted target variable, $X$ are the descriptive variables, and $A$ is the protected variable with groups ($a$). Then, selection rate is defined as the true and false positive outcomes divided by both actual positive and negative outcomes, calculating the probability of predicted ‘positive’ outcome: 

\begin{equation}
Pr(Y = 1) = \frac{TP + FP}{P + N}
\end{equation}
\bigskip

\textbf{Demographic parity} examines the selection rates of all groups in a protected variable as to whether each group is equally selected at the same rate as other groups. Specifically, we utilize demographic parity ratio between the minimum and the maximum selection rate, across all groups of the protected variable. The demographic parity ratio of $1$ means that all groups have the same selection rate. A binary classification model satisfies demographic parity if

\begin{equation}
Pr\left[\hat{Y} = 1 \mid A = a\right] = Pr\left[\hat{Y} = 1\right] \, \forall \, a
\end{equation}
\bigskip

\textbf{False positive rate difference} is the difference between the maximum and the minimum of selection rate when the fraction of data predicted as $1$, across all groups of the protected variable. False negative rate difference is the difference between the maximum and the minimum of selection rate when the fraction of data predicted as $0$, across all groups of the protected variable. 

False positive rate is defined as; $Pr\left[\hat{Y} = 1 \mid A = a, Y = 0 \right]$, and false negative rate is defined as; $Pr\left[\hat{Y} = 0 \mid A = a, Y = 1 \right]$. \bigskip

\textbf{Equalized odds} represents the probability of the predicted target value to be positive at the same selection rate for all groups in the protected variable. For a fair model, equalized odds is expected to be the same for all groups. Thus, the equalized odds difference of $0$ means that all values have the same selection rate along with the same true positive, true negative, false positive, and false negative rates. The higher equalized odds difference indicates higher bias towards the group in the protected variable with the highest equalized odds \citep{hardt2016equality,stevens2020explainability,fu2021crowds}. Equalized odds is defined as:

\begin{equation*}
Pr\left[\hat{Y} = 1 \mid A = 0, Y = y\right] = Pr\left[\hat{Y} = 1 \mid A = 1, Y = y\right], \; y \in \{0, 1\}
\end{equation*}

\clearpage

\section{Tables}
\label{sec:TablesAppendix}
\vspace{-3mm}

\subsection{Data Description for TEDS-D from SAMSHA}
\label{Appendix:sec:TablesAppendix}

\begin{table}[h!]
\centering
\begin{tabular}{p{5cm}p{10cm}}
\toprule[2pt]
\textbf{Variables} & \textbf{Description} \\ 
\midrule[1pt]
ARRESTS & Arrests in past 30 days prior to admission \\ 
\midrule[1pt]
ARRESTS\_D & Arrests in past 30 days prior to discharge \\ 
\midrule[1pt]
CBSA2010 & Core Based Statistical Area\textsuperscript{\textregistered} (CBSA) for both metro and micro areas. \\ 
\midrule[1pt]
DAYWAIT / Admission Waiting Time & Days waiting to enter substance use treatment \\ 
\midrule[1pt]
DIVISION / US Region & Census division \\ 
\midrule[1pt]
DSMCRIT / SUD Diagnosis & DSM diagnosis used to identify the substance use problem for treatment. \\ 
\midrule[1pt]
EMPLOY / Employment & Employment status at admission \\ 
\midrule[1pt]
EMPLOY\_D & Employment status at discharge \\ 
\midrule[1pt]
FREQ1 & Frequency of use at admission (primary) \\ 
\midrule[1pt]
FREQ1\_D & Frequency of use at discharge (primary) \\ 
\midrule[1pt]
FREQ\_ATND SELF\_HELP & Attendance at substance use self-help groups in past 30 days prior to admission \\ 
\midrule[1pt]
FREQ\_ATND SELF\_HELP\_D & Attendance at substance use self-help groups in past 30 days prior to discharge \\ 
\midrule[1pt]
FRSTUSE2 & Age at first use (secondary) \\ 
\midrule[1pt]
HLTHINS / Health Insurance & Health insurance \\ 
\midrule[1pt]
LIVARAG & Living arrangements at admission; homeless, a dependent or living independent \\ 
\midrule[1pt]
LIVARAG\_D / Living Arr. at Discharge & Living arrangements at discharge; homeless, a dependent or living independent \\ 
\midrule[1pt]
MARSTAT / Marital Status & Marital status \\ 
\midrule[1pt]
METHUSE / Opioid Therapy & Medication-assisted opioid therapy \\ 
\midrule[1pt]
PRIMINC / Income Source & Source of income/support \\ 
\midrule[1pt]
PRIMPAY / Payment Source & Payment source, primary (expected or actual) \\ 
\midrule[1pt]
PSOURCE & Referral source \\ 
\midrule[1pt]
PSYPROB / Mental-SUD Comorbid & Co-occurring mental and substance use disorders \\ 
\midrule[1pt]
RACE & Race \\ 
\midrule[1pt]
REASON & Reason for discharge \\ 
\midrule[1pt]
STFIPS & Census state FIPS code \\ 
\midrule[1pt]
SUB1 & Substance use at admission (primary) \\ 
\midrule[1pt]
SUB1\_D / Substance at Discharge & Substance use at discharge (primary) \\ 
\midrule[1pt]
SUB2\_D & Substance use at discharge (secondary) \\ 
\bottomrule[2pt]
\end{tabular}
\caption{Descriptive Summary for TEDS-D (Full List). The variables presented in this table were selected by our feature selection process.}
\label{Appendix:table:descriptive_full_list}
\end{table}

\subsection{Selection Rate Parity Results for Inpatient and Outpatient}
\label{Appendix:sec:SR_In-OutPatient}

\subsubsection{Inpatient Selection Rates}

\begin{table}[h!]
\centering
\begin{tabular}{p{5cm}p{2cm}p{2cm}p{2cm}}
\toprule[2pt]
\textbf{SUD Diagnosis - Inpatient} & \textbf{SR - LGBM} & \textbf{SR - RF} & \textbf{Count} \\ 
\midrule[1pt]
Alcohol-induced disorder & 0.652174 & 0.651376 & 369 \\ 
\midrule[1pt]
Substance-induced disorder & 0.696822 & 0.734177 & 1242 \\ 
\midrule[1pt]
Alcohol intoxication & 0.670391 & 0.670330 & 661 \\ 
\midrule[1pt]
Alcohol dependence & 0.721772 & 0.777070 & 52489 \\ 
\midrule[1pt]
Opioid dependence & 0.666837 & 0.715419 & 52528 \\ 
\midrule[1pt]
Cocaine dependence & 0.688514 & 0.732673 & 15973 \\ 
\midrule[1pt]
Cannabis dependence & 0.463297 & 0.481547 & 8394 \\ 
\midrule[1pt]
Other substance dependence & 0.603177 & 0.620842 & 27908 \\ 
\midrule[1pt]
Alcohol abuse & 0.498382 & 0.565916 & 2127 \\ 
\midrule[1pt]
\rowcolor[gray]{0.8} Cannabis abuse & 0.329710 & 0.465704 & 1065 \\ 
\midrule[1pt]
Other substance abuse & 0.487310 & 0.530303 & 1981 \\ 
\midrule[1pt]
Opioid abuse & 0.479208 & 0.522682 & 1745 \\ 
\midrule[1pt]
Cocaine abuse & 0.588997 & 0.653614 & 1016 \\ 
\midrule[1pt]
Anxiety disorders & 0.695652 & 0.634615 & 204 \\ 
\midrule[1pt]
Depressive disorders & 0.565217 & 0.601399 & 497 \\ 
\midrule[1pt]
Schizophrenia/other psychotic disorders & 0.459459 & 0.490909 & 218 \\ 
\midrule[1pt]
Bipolar disorders & 0.425532 & 0.631579 & 205 \\ 
\midrule[1pt]
\rowcolor[gray]{0.8} Attention Deficit/disruptive behavior disorders & 0.285714 & 0.428571 & 31 \\ 
\midrule[1pt]
Other mental health condition & 0.721341 & 0.764602 & 30019 \\ 
\midrule[1pt]
\rowcolor[gray]{0.8} Missing/unknown & 0.362580 & 0.364122 & 87270 \\ 
\bottomrule[2pt]
\end{tabular}
\caption{Selection Rate (SR) values for subgroups in substance use disorders diagnosis from DSM-5, based on LightGBM (LGBM) and Random Forest (RF) models.}
\label{Appendix:table:inpatient_sr_values-6}
\end{table}

\begin{table}[h!]
\centering
\begin{tabular}{p{6cm}p{2cm}p{2cm}p{2cm}}
\toprule[2pt]
\textbf{Substance at Discharge - Inpatient} & \textbf{SR - LGBM} & \textbf{SR - RF} & \textbf{Count} \\ 
\midrule[1pt]
None & 0.596670 & 0.621484 & 19537 \\ 
\midrule[1pt]
Alcohol & 0.633663 & 0.674214 & 85036 \\ 
\midrule[1pt]
Cocaine/crack & 0.641452 & 0.672100 & 27050 \\ 
\midrule[1pt]
\rowcolor[gray]{0.8} Marijuana/hashish & 0.420937 & 0.451407 & 17272 \\ 
\midrule[1pt]
Heroin & 0.592756 & 0.625342 & 57305 \\ 
\midrule[1pt]
Non-prescription methadone & 0.773810 & 0.760000 & 1036 \\ 
\midrule[1pt]
Other opiates and synthetics & 0.626711 & 0.668940 & 12532 \\ 
\midrule[1pt]
PCP & 0.537217 & 0.565657 & 1418 \\ 
\midrule[1pt]
Hallucinogens & 0.433121 & 0.500000 & 864 \\ 
\midrule[1pt]
Methamphetamine/speed & 0.473366 & 0.483333 & 50219 \\ 
\midrule[1pt]
Other amphetamines & 0.578606 & 0.653451 & 2169 \\ 
\midrule[1pt]
Other stimulants & 0.495652 & 0.653543 & 490 \\ 
\midrule[1pt]
Benzodiazepines & 0.683633 & 0.730769 & 3112 \\ 
\midrule[1pt]
Other tranquilizers & 0.444444 & 0.571429 & 48 \\ 
\midrule[1pt]
Barbiturates & 0.555556 & 0.650000 & 85 \\ 
\midrule[1pt]
Other sedatives or hypnotics & 0.513274 & 0.521739 & 353 \\ 
\midrule[1pt]
Inhalants & 0.509434 & 0.553191 & 166 \\ 
\midrule[1pt]
Over-the-counter medications & 0.466667 & 0.586957 & 157 \\ 
\midrule[1pt]
\rowcolor[gray]{0.8} Other drugs & 0.335404 & 0.376582 & 2065 \\ 
\midrule[1pt]
Missing/unknown/not collected/invalid & 0.511356 & 0.527628 & 5113 \\ 
\bottomrule[2pt]
\end{tabular}
\caption{Selection Rate (SR) values for the TEDS-D values in the variable for substance use at discharge, based on LightGBM (LGBM) and Random Forest (RF) models. The substances may include Alcohol, cocaine, marijuana, or other drugs.}
\label{Appendix:table:substance_discharge_inpatient-7}
\end{table}

\begin{table}[h!]
\centering
\begin{tabular}{p{5cm}p{2cm}p{2cm}p{2cm}}
\toprule[2pt]
\textbf{Race - Inpatient} & \textbf{SR - LGBM} & \textbf{SR - RF} & \textbf{Count} \\ 
\midrule[1pt]
\rowcolor[gray]{0.8} Alaska Native & 0.172249 & 0.160622 & 638 \\ 
\midrule[1pt]
\rowcolor[gray]{0.8} American Indian & 0.396648 & 0.411148 & 6431 \\ 
\midrule[1pt]
Asian/Pacific Islander & 0.250000 & 0.833333 & 502 \\ 
\midrule[1pt]
Black/African American & 0.634432 & 0.683192 & 58175 \\ 
\midrule[1pt]
\rowcolor[gray]{0.8} White & 0.586600 & 0.614980 & 184335 \\ 
\midrule[1pt]
\rowcolor[gray]{0.8} Asian & 0.321773 & 0.426070 & 3338 \\ 
\midrule[1pt]
\rowcolor[gray]{0.8} Other Single Race & 0.531380 & 0.553687 & 21901 \\ 
\midrule[1pt]
\rowcolor[gray]{0.8} Two or more Races & 0.402254 & 0.425249 & 5585 \\ 
\midrule[1pt]
\rowcolor[gray]{0.8} Native Hawaiian Other Pacific Islander & 0.225806 & 0.239482 & 1912 \\ 
\midrule[1pt]
\rowcolor[gray]{0.8} Unknown/Not collected & 0.603870 & 0.597970 & 3190 \\ 
\bottomrule[2pt]
\end{tabular}
\caption{Selection Rate (SR) values for groups in Race, based on LightGBM (LGBM) and Random Forest (RF) models.}
\label{Appendix:table:race_inpatient-8}
\end{table}

\begin{table}[h!]
\centering
\begin{tabular}{p{5cm}p{2cm}p{2cm}p{2cm}}
\toprule[2pt]
\textbf{Wait-time for Admission - Inpatient} & \textbf{SR - LGBM} & \textbf{SR - RF} & \textbf{Count} \\ 
\midrule[1pt]
0 & 0.568401 & 0.598761 & 97377 \\ 
\midrule[1pt]
1–7 & 0.590378 & 0.627231 & 37542 \\ 
\midrule[1pt]
\rowcolor[gray]{0.8} 8–14 & 0.378011 & 0.389876 & 9198 \\ 
\midrule[1pt]
\rowcolor[gray]{0.8} 15–30 & 0.316186 & 0.328947 & 7915 \\ 
\midrule[1pt]
\rowcolor[gray]{0.8} 31 or more & 0.251914 & 0.256804 & 4647 \\ 
\midrule[1pt]
Missing/unknown/not collected/invalid & 0.623153 & 0.657283 & 129157 \\ 
\bottomrule[2pt]
\end{tabular}
\caption{Selection Rate (SR) values for number of days patients wait to be admitted, based on LightGBM (LGBM) and Random Forest (RF) models.}
\label{Appendix:table:wait_time_admission-9}
\end{table}

\begin{table}[h!]
\centering
\begin{tabular}{p{5cm}p{2cm}p{2cm}p{2cm}}
\toprule[2pt]
\textbf{Marital Status - Inpatient} & \textbf{SR - LGBM} & \textbf{SR - RF} & \textbf{Count} \\ 
\midrule[1pt]
Never married & 0.595717 & 0.632822 & 165621 \\ 
\midrule[1pt]
Now married & 0.704927 & 0.732929 & 25420 \\ 
\midrule[1pt]
Separated & 0.629059 & 0.670941 & 17077 \\ 
\midrule[1pt]
Divorced, widowed & 0.665436 & 0.687403 & 36747 \\ 
\midrule[1pt]
\rowcolor[gray]{0.8} Missing/unknown/not collected/invalid & 0.315293 & 0.329429 & 40996 \\ 
\bottomrule[2pt]
\end{tabular}
\caption{Selection Rate (SR) values for marital status of inpatients, based on LightGBM (LGBM) and Random Forest (RF) models.}
\label{Appendix:table:marital_status_inpatient-10}
\end{table}

\begin{table}[h!]
\centering
\begin{tabular}{p{5cm}p{2cm}p{2cm}p{2cm}}
\toprule[2pt]
\textbf{Living Arrangement at Discharge - Inpatient} & \textbf{SR - LGBM} & \textbf{SR - RF} & \textbf{Count} \\ 
\midrule[1pt]
Homeless & 0.672670 & 0.727102 & 45959 \\ 
\midrule[1pt]
\rowcolor[gray]{0.8} Dependent living & 0.438125 & 0.465252 & 94538 \\ 
\midrule[1pt]
Independent living & 0.676709 & 0.703012 & 120858 \\ 
\midrule[1pt]
\rowcolor[gray]{0.8} Missing/unknown/not collected/invalid & 0.389796 & 0.414694 & 24337 \\ 
\bottomrule[2pt]
\end{tabular}
\caption{Selection Rate (SR) values for living arrangements of patients at discharge, based on LightGBM (LGBM) and Random Forest (RF) models.}
\label{Appendix:table:living_arrangement_discharge-11}
\end{table}

\begin{table}[h!]
\centering
\begin{tabular}{p{4.5cm}p{2.5cm}p{2.5cm}p{2.5cm}}
\toprule[2pt]
\textbf{U.S. Region - Inpatient} & \textbf{SR - LGBM} & \textbf{SR - RF} & \textbf{Count} \\ 
\midrule[1pt]
\rowcolor[gray]{0.8} U.S. territories & 0.457447 & 0.440860 & 352 \\ 
\midrule[1pt]
\rowcolor[gray]{0.8} New England & 0.397256 & 0.401022 & 25848 \\ 
\midrule[1pt]
Middle Atlantic & 0.723061 & 0.765022 & 80328 \\ 
\midrule[1pt]
East North Central & 0.755494 & 0.851073 & 30814 \\ 
\midrule[1pt]
\rowcolor[gray]{0.8} West North Central & 0.499433 & 0.516481 & 36983 \\ 
\midrule[1pt]
\rowcolor[gray]{0.8} South Atlantic & 0.620021 & 0.666367 & 26468 \\ 
\midrule[1pt]
East South Central & 0.707609 & 0.717105 & 15352 \\ 
\midrule[1pt]
\rowcolor[gray]{0.8} West South Central & 0.594198 & 0.621959 & 21397 \\ 
\midrule[1pt]
\rowcolor[gray]{0.8} Mountain & 0.443379 & 0.430439 & 13584 \\ 
\midrule[1pt]
\rowcolor[gray]{0.8} Pacific & 0.212036 & 0.212346 & 34992 \\ 
\bottomrule[2pt]
\end{tabular}
\caption{Selection Rate (SR) values for regions of the US with census categories, based on LightGBM (LGBM) and Random Forest (RF) models.}
\label{Appendix:table:us_region_inpatient-12}
\end{table}

\clearpage

\subsubsection{Outpatient Selection Rates}

\begin{table}[h!]
\centering
\begin{tabular}{p{5.5cm}p{2cm}p{2cm}p{2cm}}
\toprule[2pt]
\textbf{Substance at Discharge - Outpatient} & \textbf{SR - LGBM} & \textbf{SR - RF} & \textbf{Count} \\ 
\midrule[1pt]
\rowcolor[gray]{0.8} None & 0.469740 & 0.546303 & 70948 \\ 
\midrule[1pt]
\rowcolor[gray]{0.8} Alcohol & 0.537184 & 0.618798 & 278153 \\ 
\midrule[1pt]
Cocaine/crack & 0.631623 & 0.684367 & 61897 \\ 
\midrule[1pt]
\rowcolor[gray]{0.8} Marijuana/hashish & 0.497607 & 0.585360 & 160470 \\ 
\midrule[1pt]
\rowcolor[gray]{0.8} Heroin & 0.529221 & 0.594638 & 204759 \\ 
\midrule[1pt]
\rowcolor[gray]{0.8} Non-prescription methadone & 0.502088 & 0.609603 & 6857 \\ 
\midrule[1pt]
\rowcolor[gray]{0.8} Other opiates and synthetics & 0.604962 & 0.653853 & 88721 \\ 
\midrule[1pt]
\rowcolor[gray]{0.8} PCP & 0.595376 & 0.671965 & 3476 \\ 
\midrule[1pt]
\rowcolor[gray]{0.8} Hallucinogens & 0.485804 & 0.646688 & 2321 \\ 
\midrule[1pt]
Methamphetamine/speed & 0.632631 & 0.691967 & 126963 \\ 
\midrule[1pt]
\rowcolor[gray]{0.8} Other amphetamines & 0.580948 & 0.627082 & 7620 \\ 
\midrule[1pt]
Other stimulants & 0.702557 & 0.761777 & 2562 \\ 
\midrule[1pt]
\rowcolor[gray]{0.8} Benzodiazepines & 0.624947 & 0.661858 & 7503 \\ 
\midrule[1pt]
\rowcolor[gray]{0.8} Other tranquilizers & 0.413793 & 0.551724 & 164 \\ 
\midrule[1pt]
\rowcolor[gray]{0.8} Barbiturates & 0.171975 & 0.191083 & 721 \\ 
\midrule[1pt]
\rowcolor[gray]{0.8} Other sedatives or hypnotics & 0.616740 & 0.678414 & 1482 \\ 
\midrule[1pt]
\rowcolor[gray]{0.8} Inhalants & 0.593548 & 0.645161 & 568 \\ 
\midrule[1pt]
\rowcolor[gray]{0.8} Over-the-counter medications & 0.546392 & 0.577320 & 396 \\ 
\midrule[1pt]
Other drugs & 0.763027 & 0.798077 & 10267 \\ 
\midrule[1pt]
Missing/unknown/not collected/invalid & 0.826092 & 0.857653 & 134080 \\ 
\bottomrule[2pt]
\end{tabular}
\caption{Selection Rate (SR) values for substance use at discharge for outpatient, based on LightGBM (LGBM) and Random Forest (RF) models.}
\label{Appendix:table:substance_discharge_outpatient-13}
\end{table}

\begin{table}[h!]
\centering
\begin{tabular}{p{5.5cm}p{2.5cm}p{2.5cm}p{2cm}}
\toprule[2pt]
\textbf{Race - Outpatient} & \textbf{SR - LGBM} & \textbf{SR - RF} & \textbf{Count} \\ 
\midrule[1pt]
\rowcolor[gray]{0.8} Alaska Native & 0.299233 & 0.372123 & 2629 \\ 
\midrule[1pt]
American Indian & 0.778890 & 0.816750 & 30823 \\ 
\midrule[1pt]
Asian/Pacific Islander & 0.574627 & 0.753731 & 1609 \\ 
\midrule[1pt]
\rowcolor[gray]{0.8} Black/African American & 0.517813 & 0.611038 & 220283 \\ 
\midrule[1pt]
\rowcolor[gray]{0.8} White & 0.594243 & 0.655253 & 755143 \\ 
\midrule[1pt]
\rowcolor[gray]{0.8} Asian & 0.387506 & 0.550493 & 12529 \\ 
\midrule[1pt]
\rowcolor[gray]{0.8} Other Single Race & 0.497494 & 0.563791 & 76859 \\ 
\midrule[1pt]
\rowcolor[gray]{0.8} Two or more Races & 0.541241 & 0.596620 & 18717 \\ 
\midrule[1pt]
\rowcolor[gray]{0.8} Native Hawaiian Other Pacific Islander & 0.596491 & 0.647528 & 5712 \\ 
\midrule[1pt]
Unknown/Not collected & 0.813725 & 0.844926 & 45548 \\ 
\bottomrule[2pt]
\end{tabular}
\caption{Selection Rate (SR) values on race subgroup for outpatient, based on LightGBM (LGBM) and Random Forest (RF) models.}
\label{Appendix:table:race_outpatient-14}
\end{table}

\begin{table}[h!]
\centering
\begin{tabular}{p{5.5cm}p{2.5cm}p{2.5cm}p{2cm}}
\toprule[2pt]
\textbf{U.S. Region - Outpatient} & \textbf{SR - LGBM} & \textbf{SR - RF} & \textbf{Count} \\ 
\midrule[1pt]
\rowcolor[gray]{0.8} U.S. territories & 0.303390 & 0.466102 & 2023 \\ 
\midrule[1pt]
\rowcolor[gray]{0.8} New England & 0.493094 & 0.571869 & 76336 \\ 
\midrule[1pt]
\rowcolor[gray]{0.8} Middle Atlantic & 0.476601 & 0.574001 & 214779 \\ 
\midrule[1pt]
\rowcolor[gray]{0.8} East North Central & 0.522693 & 0.630076 & 95142 \\ 
\midrule[1pt]
\rowcolor[gray]{0.8} West North Central & 0.496673 & 0.601275 & 111527 \\ 
\midrule[1pt]
\rowcolor[gray]{0.8} South Atlantic & 0.634243 & 0.670520 & 272440 \\ 
\midrule[1pt]
East South Central & 0.863360 & 0.872863 & 73792 \\ 
\midrule[1pt]
\rowcolor[gray]{0.8} West South Central & 0.282824 & 0.387423 & 46632 \\ 
\midrule[1pt]
Mountain & 0.773046 & 0.804573 & 172136 \\ 
\midrule[1pt]
\rowcolor[gray]{0.8} Pacific & 0.490297 & 0.566029 & 105512 \\ 
\bottomrule[2pt]
\end{tabular}
\caption{Selection Rate (SR) values on US regions according to the census for outpatient, based on LightGBM (LGBM) and Random Forest (RF) models.}
\label{Appendix:table:us_region_outpatient-15}
\end{table}

\begin{table}[h!]
\centering
\begin{tabular}{p{5.5cm}p{2.5cm}p{2.5cm}p{2cm}}
\toprule[2pt]
\textbf{Payment Source - Outpatient} & \textbf{SR - LGBM} & \textbf{SR - RF} & \textbf{Count} \\ 
\midrule[1pt]
\rowcolor[gray]{0.8} Self-pay & 0.412974 & 0.504767 & 36829 \\ 
\midrule[1pt]
Private insurance & 0.677791 & 0.720187 & 31077 \\ 
\midrule[1pt]
Medicare & 0.822420 & 0.880854 & 18477 \\ 
\midrule[1pt]
Medicaid & 0.741308 & 0.790917 & 291173 \\ 
\midrule[1pt]
\rowcolor[gray]{0.8} Other government payments & 0.516697 & 0.604019 & 109193 \\ 
\midrule[1pt]
\rowcolor[gray]{0.8} No charge & 0.247555 & 0.384257 & 16312 \\ 
\midrule[1pt]
\rowcolor[gray]{0.8} Other & 0.421728 & 0.464470 & 36293 \\ 
\midrule[1pt]
\rowcolor[gray]{0.8} Missing/unknown/not collected/invalid & 0.540172 & 0.609355 & 630443 \\ 
\bottomrule[2pt]
\end{tabular}
\caption{Selection Rate (SR) values on primary payer for outpatient, based on LightGBM (LGBM) and Random Forest (RF) models.}
\label{Appendix:table:payment_source_outpatient-16}
\end{table}

\end{appendices}